%% file: main.tex
\documentclass[12pt]{article}

\usepackage[utf8]{inputenc} 
\usepackage{parskip} 
\usepackage[a4paper, margin=1in]{geometry}  
\usepackage{graphicx} 
\usepackage{rotating}
\usepackage{subcaption} 
\usepackage{float} 
\usepackage{booktabs} 
\usepackage{multirow}
\usepackage{multicol}
\usepackage{enumitem} 
\usepackage{hyperref} 
\hypersetup{
    colorlinks=true,   
    linkcolor=blue,    
    citecolor=blue,    
    urlcolor=blue,     
    linkbordercolor={0 0 0},  
    pdfborder={0 0 0}   
}
\usepackage{amsmath, amssymb, amsthm}  
\usepackage{mathtools}
\usepackage{siunitx} 
\usepackage{bm} 
\usepackage{algorithm, algpseudocode} 
\usepackage{natbib}
\algrenewcommand\algorithmicrequire{\textbf{Input:}}
\algrenewcommand\algorithmicensure{\textbf{Output:}}
\usepackage{csquotes} 

\theoremstyle{definition}

\newcommand{\norm}[1]{\left\lVert#1\right\rVert}
\newcommand{\absnorm}[1]{\left|#1\right|}

\usepackage{authblk} 

\begin{document}

\title{A functional mixed-effects model for the joint analysis of amplitude and phase variations in function-valued traits}
\author{Yilin Chen}
\author{Davide Pigoli}
\affil{Department of Mathematics, King's College London}
\author{John A.D. Aston}
\affil{Department of Pure Mathematics and Mathematical Statistics, University of Cambridge}
\author{Patrick A. Carter}
\affil{School of Biological Sciences, Washington State University}
\date{\today}

\maketitle

\begin{abstract}
    Quantitative genetics is the study of continuous traits and the statistical analysis of phenotypic variation, that is, the observable differences in physical traits among individuals within a population. If family relationships between individuals are known, phenotypic variation can be partitioned into genetic and environmental components. Some traits, such as growth trajectories, can be described as a function of a continuous index and are known as function-valued traits. However, function-valued traits from different individuals often vary not only in their shape (amplitude) but also in the speed at which the biological process develops (phase). In this work, we present and compare two approaches to account for amplitude and phase variations in estimating the additive genetic and environmental components. The first one follows from a conventional framework where warping functions remove phase variations, and then the aligned curves are analysed. The second approach adopts the perspective of looking at the aligned curve and the warping function as a joint functional object. In both cases, we propose a functional mixed-effects framework that uses functional principal components as the data-driven basis for genetically correlated data. The proposed procedure is applied to simulations and a real-case study of \textit{Tribolium castaneum} growth trajectories.
\end{abstract}


\input{1_Introduction}
\input{2_GeneticFMEM}

\input{3_JointFMEM}

\input{4_Simulation}
\input{5_RealData}

\input{6_Discussion}

\section*{Acknowledgements}
    The first author gratefully acknowledges funding support from the Kings-China Scholarship Council Scholarship Programme (File No.~202308060098). Computational support for simulation and bootstrap studies was provided by CREATE \citep{kingscollegelondon_2022_}.
    
\section*{Data availability}
The \emph{Tribolium castaneum} growth dataset was originally published with \citet{irwin_2013_constraints} and deposited at Dryad: doi:\href{https://datadryad.org/dataset/doi:10.5061/dryad.61t9h}{10.5061/dryad.61t9h}. The code to reproduce the results in the simulation study and real data analysis is available at
\\
\href{https://github.com/yilinchen98/FMEMs-quantitative-genetics}{https://github.com/yilinchen98/FMEMs-quantitative-genetics}.

\bibliographystyle{abbrvnat}
\setcitestyle{authoryear,open={((},close={))}} 

\bibliography{references}

\end{document}

%% file: 1_Introduction.tex
\section{Introduction}\label{Sec:Introduction}
Variation in phenotypes, as observable differences in traits (phenotypes) among individuals within a population, is critically important to the ecology and evolution of species. Many features of an individual may vary continuously in some manner. A phenotypic trait whose value is a function of some continuous index, such as time, is known as a function-valued (FV) trait \citep{kingsolver_2001_variation}. For example, growth curves of body mass measured over time are widely studied as they provide valuable insights into evolutionary fitness \citep{cheverud_1983_quantitative,mousseau_1987_natural}. Such traits, by nature, can be assessed for continuous genetic variation using a quantitative genetic approach.

Quantitative genetics is the study of continuous traits and the statistical analysis of the relative contributions of genetic and environmental effects to phenotypic variation. If family relationships within a population are measured and known, then phenotypic variation can be partitioned into genetic and environmental components \citep{lynch_1998_genetics,falconer1996pres}. Furthermore, if family relationships within the population are sufficiently detailed, the genetic component of variance can be decomposed into specific genetic effects, such as additive and dominance variance. Additive genetic variation is essentially the effect of allelic substitutions and so is the component of genetic variation that is passed directly from parents to offspring, and so is the component of genetic variation of most interest to evolutionary biologists and plant and animal breeders. The ratio between additive genetic variation and total phenotypic variation is defined as the heritability of the trait. A higher heritability indicates that a larger proportion of the phenotypic variation is due to genetic factors; a non-zero heritability is required for a trait to be able to respond to selection. Thus, research often focuses on estimating additive genetic variation, which provides insights into the extent to which genetic factors play a role in trait variability and evolution within a population.

FV traits are naturally described as mathematical functions. Their functional nature suggests this field could benefit from functional data analysis (FDA), which offers methods for analysing continuously varying data. One major challenge in analysing functional data is that they often present amplitude and phase variations. In growth curves, amplitude variation refers to differences in the magnitude of growth, such as the maximum body mass of each individual. Phase variation, on the other hand, captures differences in the timing of growth spurts. This issue, also known as curve misalignment, can occur due to measurement uncertainties or because the rigid metric of physical time may not be directly relevant to the internal dynamics of many biological systems \citep{ramsay_2005_functional}. Confounding these two types of variations makes it difficult to accurately compare growth patterns across different individuals or analyse the underlying dynamics. Standard statistical methods assume that the same argument $t$ corresponds across all functions. However, it is possible that the time axis itself is subject to noise in a way that preserves the ordering of observations \citep{ebert_2021_curve}. Depending on the context, it is often of interest to either (i) align functions before analysis (of amplitude only) if phase variation is not of interest, which is also known as registration, or (ii) to separate amplitude and phase variation and analyse them separately or jointly \citep{gertheiss_2024_functional}.

Curve registration aligns curves via transforming time with warping functions and provides a solution to separate amplitude and phase variations. \citet{gasser_1995_searching} discussed the structural analysis of curve samples. They defined structure points/landmarks as the local extrema and inflexion points. \citet{wang_1997_alignment} proposed a dynamic time-warping approach to align or shift curves so that the landmarks occur at the same transformed time, with smooth interpolation applied between the landmarks. There are two drawbacks to this approach. Structural points/landmarks may be missing from certain curves, and estimating landmark timing can be difficult. Continuous registration was proposed as an alternative to landmark registration. A classical framework minimises the loss function between one function and a second aligned function over time warping. However, the $L^2$ metric as the loss function is disappointing if a flexible warping function is applied, resulting in significantly distorted amplitude characteristics \citep{marron_2015_functional}. This is called the pinch effect, and it happens as the least squares criterion forces warping to minimise amplitude difference by squeezing out of the region where amplitudes differ \citep{ramsay_2005_functional}. The pinch effect can be mitigated by regularisation. \citet{ramsay_1998_curve} extended the global fitting criterion proposed by  \citet{silverman_1995_incorporating} and proposed a flexible family of smooth monotone warping functions incorporating a penalised term. A novel framework of shape-preserving registration using the Fisher-Raw metric was proposed by \citet{srivastava_2011_registration}. Transformation of the Fisher-Raw metric to the standard $L^2$ metric was done via a square-root velocity function (SRVF). This provided a consistent estimator for signals observed under random warping and scaling. Registering curves by simply removing phase variation can give problematic results if phase variation is of research interest or the amplitude and phase variations are correlated. Recent studies have shifted to a joint analysis of the two sources of variation. In particular, functional principal component analysis (FPCA) has been widely used to extract a small number of functional principal components (FPCs) that summarise the coupled variation. \citet{lee_2016_combined} proposed a functional combined principal component analysis (functional combined PCA) to jointly analyse the amplitude and phase functions by concatenating them with a scaling parameter. The nonlinear combined FPCs simultaneously capture the amplitude, phase and their association. \citet{happ_2019_a} embedded the functional combined FPA approach into a general framework for multivariate FPCA of the two types of variations. An alternative view was introduced by \citet{hadjipantelis_2015_unifying}. They discussed a two-step approach to unify amplitude and phase analysis by applying FPCA to both functions, followed by a linear mixed-effects model on individual principal component scores to explore their relationship.

Once we have separated amplitude and phase variation, the analysis in genetic modelling can focus on understanding the genetic and environmental components of phenotypic variation. This allows important biological questions to be addressed, in particular, identifying possible evolutionary constraints by estimating the additive genetic covariance function of FV traits. Numerous studies have explored this topic, providing a foundation for further investigation. \citet{kirkpatrick_1989_a} introduced a quantitative approach to estimate the genetic covariance function (CF) in infinite-dimensional traits using orthogonal function expansion on the estimated genetic covariance matrix. Later, \citet{kirkpatrick_1990_analysis} developed two models to estimate genetic covariance functions: a ``full" model that fully fits the data, and a ``reduced" model that produces a smoothed estimate that takes into account sampling errors in the data using weighted least square methods. An alternative approach using restricted maximum likelihood (REML) estimation was discussed in \citet{meyer_1997_estimation} and \citet{meyer_1998_estimating}, where a random regression (RR) model was applied to estimate covariance functions directly from data. In particular, \citet{meyer_1998_estimating} showed that the CF and RR models are mathematically equivalent, and the latter does not require computing a covariance matrix. Random regression or mixed-effects models are well-suited for capturing both population-level trends and individual-level variation and are broadly applied in analysing longitudinal data. An extension of classical mixed-effects models to a functional framework introduces the functional mixed-effects models (FMEMs), where both fixed and random effects are modelled by nonparametric curves through basis expansion. The term FMEM was first introduced by \citet{guo_2002_functional}, in which cubic splines were used as the basis functions. Subsequent research generalised this concept to a variety of model structures. \citet{morris_2006_waveletbased} expanded this concept to wavelet-based models that incorporate multiple levels of random-effect functions and allow for curve-to-curve deviations. Unlike using data-agnostic basis functions, FPCA plays a role in dimension reduction, and FPCs provide data-driven basis functions for model fitting. FPC-based models were studied by \citet{greven_2010_longitudinal} and \citet{cederbaum_2015_functional}. 

In this paper, we investigate the decomposition of phenotypic variation into genetic and environmental components using a mixed-effects framework tailored to function-valued genetic data, and we propose two approaches to account for amplitude and phase variability. The first approach removes phase variations in the data pre-processing stage, and the subsequent analysis is on the aligned curves. The second approach embeds the combined FPCs discussed in \citet{lee_2016_combined} in our mixed-effect model and simultaneously captures the covariations of amplitude and phase functions. The rest of this paper is organised as follows. Section~\ref{Sec:GFMEM} and Section~\ref{Sec:Joint} introduce the model for aligned curves and the joint model of amplitude and phase variations, respectively. Simulation studies and model performance are presented in Section~\ref{Sec:Simulation}. In Section~\ref{Sec:TCData}, we apply both models to the dataset of red flour beetle growth curves. Finally, Section~\ref{Sec:Discussion} discusses some limitations and outlines potential directions for future work. 

%% file: 2_GeneticFMEM.tex
\section{Genetic functional mixed-effects model}\label{Sec:GFMEM}
\subsection{Quantitative genetics for function-valued traits}
As mentioned in the previous section, quantitative genetics deals with biological traits characterised by continuous variation in a population. A key idea is that phenotypic (i.e. observable) variation can be decomposed into additive genetic effect and environmental effect. For a function-valued trait $y(t)$, this means that its (phenotypic) covariance function can be written as
\begin{equation*}
    cov(y(t),y(s)) = P(t,s) = G(t,s) + E(t,s), 
\end{equation*}
where $G(t,s)$ denotes the additive genetic covariance function and $E(t,s)$ the environmental covariance function.

Compared with a classical multivariate case, which produces a G-matrix of discrete additive genetic (co)variances, functional models allow the estimation of a G-function that describes the genetic variations between any two points on a continuous index. Once the G-function is available, the short-term evolutionary trajectory of the trait given selection gradient functions $\rho(s)$ can be estimated using the functional breeder's equation \citep{kirkpatrick_1989_a}:
$$ \Delta \bar{y}(t) = \int_\mathbb{R} G(t,s)\rho(s) \: ds.$$
Through spectral analysis of the additive genetic covariance function, it is not difficult to show that the maximum response to selection corresponds to its first eigenfunction (with the largest eigenvalue). On the other hand, the selection gradient collinear with eigenfunctions with small eigenvalues will result in minor or no change in the phenotypic variation, which is a very important type of evolutionary constraint.

Therefore, the primary goal for the analysis is to estimate the additive genetic covariance function. Several methods have been proposed for this purpose. Our approach is based on the random regression model introduced by \citet{meyer_1998_estimating}, but it incorporates FDA methodologies from two perspectives: (i) we consider amplitude and phase variations in the misaligned curves, and we then extend our model to incorporate warping functions as well as the aligned curves; (ii) we choose FPCs as data-driven basis functions to reduce the model complexity while effectively capturing its underlying structure. Numerical computation is done using the R package \textbf{lme4} \citep{bates_2015_fitting}, which fits linear mixed-effects models with flexible grouping structures, correlated and uncorrelated random effects within levels. However, \textbf{lme4} cannot be directly used to model genetic data since it assumes independence between individuals; thus, the genetic relationships among individuals from shared ancestry introduce correlations that must be incorporated in the model. To address this issue, we apply the model reparameterisation approach introduced in \citet{Harville1990} and \citet{vazquez_2010_technical}, and details of the estimation procedure are explained in the rest of this section.

\subsection{Model setup}\label{Subsec:alignedmodel}
Functional data are assumed to be realisations of a stochastic process $X(\cdot)$ that is in the $L^2$ space defined on a compact interval $\mathcal{T}$ with mean function $\mu(t) = \mathbb{E}(X(t))$ and covariance function $\Sigma(t,s) = cov(X(t), X(s))$. Without loss of generality, $\mathcal{T}$ can be taken as the unit interval $[0,1]$. In practice, we observe sample data for the \textit{i}th subject at time points $0 = t_{i1}, \dots, t_{ij} = 1$ with the corresponding measurements as $\bm{x_i} = [x_{i1}, \dots, x_{ij}]^\intercal$, where $x_{ij} = X_i(t_{ij})$. For simplicity, we assume the same sampling schedule is applied to all subjects, i.e., $t_{ij} = t_j$. In addition, we assume random measurement errors $e_{ij}$ with $\mathbb{E}(e_{ij}) = 0$ and $var(e_{ij}) = \sigma^2_{ij}$ are observed in the realisation process. Therefore, the actual measurement is $y_{ij} = x_{ij} + e_{ij}$. We then look for warping functions to align curves via transforming time to separate amplitude and phase variations. A time-warping function is an orientation-preserving diffeomorphism of the unit time interval $[0,1]$ and lies in the set
\begin{equation*}
    H = \{h: [0,1] \rightarrow [0,1] \,|\, h(0) = 0, h(1) = 1\}.
\end{equation*} 
Here, we use the continuous registration based on the Fisher-Rao metric proposed by \citet{srivastava_2011_registration}. To apply to FV traits, we assume $y_i(h(t_j)) = \tilde{y}_{ij}$ represents the registered value of a trait recorded at time $t_j$ for the $i$th subject. 

The decomposition of phenotypic variation into additive genetic and environmental components suggests a mixed-effects model with two types of random effects. Following the random regression in \citet{meyer_1998_estimating}, we introduce the genetic FMEM represented using a finite number of basis functions $\phi(t)$:
\begin{equation}\label{eqn:GFMEM1}
   \underbrace{\tilde{y}_{ij}}_{\text{registered trait}} = \underbrace{\sum_{k = 1}^{K_f} \beta_k\phi_k(t_{ij})}_{\text{fixed effect: population mean}} + \underbrace{\sum_{k=1}^{K_g} \alpha_{ik}\phi_k(t_{ij})}_{\text{additive genetic random effect}} + \underbrace{\sum_{k=1}^{K_e} \gamma_{ik}\phi_k(t_{ij})}_{\text{environmental random effect}} + \underbrace{\epsilon_{ij},}_{\text{error}} 
\end{equation}
where $\beta_k$ are the fixed-effect coefficients; $\alpha_{ik}$ and $\gamma_{ik}$ represent the $k$th additive genetic and environmental effect coefficients for the $i$th subject.

Notice that Equation~(\ref{eqn:GFMEM1}) represents a truncated form of infinite-dimensional functional data. In this work, we use FPCA to derive FPCs as the basis functions for both fixed and random effects. The critical distinction between FPCA and dimension reduction through other basis functions, such as splines, is that FPCs extract most of the variation of the functional process in $L^2$ sense \citep{wang_2016_functional}. We, therefore, reduce the model complexity to a finite (usually low) dimension and obtain data-driven basis functions to fit our model.

The covariance between any two measured points of the same subject is 
\begin{multline}\label{eqn:phencov}
    cov(\tilde{y}_{ij}, \tilde{y}_{ij'}) = \sum_{k = 1}^{K_g} \sum_{l=1}^{K_g} \phi_k(t_{ij})\phi_l(t_{ij'})\,cov(\alpha_{ik}, \alpha_{il}) \\
    + \sum_{k = 1}^{K_e} \sum_{l=1}^{K_e} \phi_k(t_{ij})\phi_l(t_{ij'})\,cov(\gamma_{ik}, \gamma_{il}) + cov(\epsilon_{ij},\epsilon_{ij'}).
\end{multline}
Estimating the additive genetic covariance function involves estimating the covariance of additive genetic coefficients over basis functions. This is analogous to estimating the covariance in a classical, finite-dimensional multivariate analysis, where maximum likelihood estimation can be applied.

Assume there are $N$ subjects $\bm{y_1}, \dots, \bm{y_N}$, each of which has $n$ observations. Considering all subjects, the vectorised response is $\bm{\tilde{y}} = [\bm{\tilde{y}_1}^\intercal, \cdots, \bm{\tilde{y}_N}^\intercal]^\intercal \in \mathbb{R}^{Nn}$. Equation~(\ref{eqn:GFMEM1}) can be written in matrix form,
\begin{equation}\label{eqn:GFMEM3}
    \bm{\tilde{y}} = \bm{X\beta + Z^G \alpha + Z^E\gamma + \epsilon}, 
\end{equation}
with $\bm{\beta}$ the vector of fixed effects, $\bm{\alpha} = [\bm{\alpha_1}^\intercal, \dots, \bm{\alpha_N}^\intercal]^\intercal \in \mathbb{R}^{N K_g}$ the vector of additive genetic coefficients, $\bm{\gamma} = [\bm{\gamma_1}^\intercal, \dots, \bm{\gamma_N}^\intercal]^\intercal \in \mathbb{R}^{N K_e}$ the vector of environmental coefficients, $\bm{\epsilon}$ vector of measurement errors. The matrices $\bm{X}$, $\bm{Z^G}$ and $\bm{Z^E}$ are the corresponding design matrices. 

The random terms are assumed to have the following distributions:
\begin{gather}\label{eqn:re}
    \begin{aligned}
        \bm{\alpha} & \sim \mathcal{N}(\bm{0}, \bm{A} \otimes \bm{C^G}) \\
        \bm{\gamma} & \sim \mathcal{N}(\bm{0}, \bm{I_N} \otimes \bm{C^E}) \\
        \bm{\epsilon} & \sim \mathcal{N}(\bm{0}, \sigma^2  \bm{I_{Nn}})
    \end{aligned}
\end{gather} 
where $\bm{C^G}$ and $\bm{C^E}$ are the additive genetic and environmental effects covariance matrices; $\bm{A}$ is the relationship matrix, which can be used to estimate the additive effects. It is derived from the family relationship of a breeding population and quantifies the genetic relatedness among individuals. Specifically, $\bm{A}$ is a positive definite matrix with each term $0 \leq A_{ij} \leq 1$ representing how closely related individuals $i$ and $j$ are. The closer the entry is to 1, the more similar the two individuals are. For instance, $A_{ij} = 0.5$ means the two individuals have both sire and dam in common and $A_{ij} = 0.25$ means the two individuals have either a sire or a dam in common. The Kronecker product ($\otimes$) is used to generate the full covariance structure for random effects.

\subsection{Details of the estimation procedure}\label{Subsec:estalign}
Model~(\ref{eqn:GFMEM3}) cannot be directly estimated using standard tools for mixed-effects models, such as the \textbf{lme4} package in R that provides functions to fit linear mixed-effects models using the penalised least squares (PLS) method. In such implementations,  model parameters are estimated using the maximum likelihood (ML) or restricted maximum likelihood (REML) criterion \cite{bates_2015_fitting}. Nonetheless, they do not allow for between-level correlations of random effects, so it is not directly applicable to genetic data. However, it is possible to reformulate Model~(\ref{eqn:GFMEM3}) so that it can be fitted within the \textbf{lme4} framework. Here, we revisit the approach proposed by  \citet{Harville1990} and \citet{vazquez_2010_technical} for multivariate genetic mixed effects models: the correlation between individuals is induced by post-multiplying the model design matrix by the Cholesky factor of the relationship matrix. In this section, we illustrate how this approach can be applied to the proposed model.

\subsubsection{Model reparameterisation}\label{Subsubsec:repara}
It is worth describing the structure of the design matrices $\bm{X}$ and $\bm{Z}$ used in the functional framework before explaining how to incorporate the additive genetic relationship into the estimation procedure. Matrix $\bm{X}$ is of dimension $Nn \times K$ matrix where each column consists of the basis function $\phi_{k}(t)$ evaluated at sampling points, i.e., 
$$
\bm{X} = 
\begin{bmatrix}
    \phi_1(t_{11}) & \dots & \phi_k(t_{11}) \\
    \vdots &  & \vdots \\
    \phi_1(t_{1n}) & \dots & \phi_k(t_{1n}) \\
    \vdots &  & \vdots \\
    \phi_1(t_{Nn}) & \dots  & \phi_k(t_{Nn})
\end{bmatrix}.
$$

We need to represent the grouping structures in mixed-effects models in a mathematical format. Let $l$ be the number of grouping levels in a mixed-effects model. In a genetic mixed-effects model with groups at the individual level, we have $l = N$. Define $\bm{J}$ as a $Nn \times N$ matrix of indicator columns for grouping factors:
$$ \bm{J} = 
\begin{bmatrix}
    1 & 0 & \dots \\
    1 & 0 & \\
    \vdots & \vdots \\
    0 & 1 & \\
    0 & 1 & \\
    \vdots & \vdots 
\end{bmatrix}.
$$
Each element $J_{hi}$ equals 1 if the observation $h$ in the dataset belongs to the $i$th subject. 

Now, we can define the random-effect design matrix $\bm{Z}$ as 
$$ \bm{Z} = (\bm{J}^\intercal * \bm{X}^\intercal)^\intercal =
\begin{bmatrix}
    \bm{J_1}^\intercal \otimes \bm{X_1}^\intercal \\
    \bm{J_2}^\intercal \otimes \bm{X_2}^\intercal \\
    \vdots \\
    \bm{J_N}^\intercal \otimes \bm{X_N}^\intercal
\end{bmatrix}
$$
where $*$ and $\otimes$ are the Khatri-Rao product \cite{g_1968_solutions} and Kronecker product; $\bm{J_i}^\intercal$ and $\bm{X_i}^\intercal$ are the $i$th row vectors of $\bm{J}$ and $\bm{X}$.

To capture genetic correlations between individuals, we first take the Cholesky decomposition of $\bm{A} = \bm{LL}^\intercal$, and define $\bm{M} = \bm{L} \otimes \bm{I}_K$, where $K$ represents the number of basis functions. The modified genetic random-effect design matrix is:
\begin{equation}
  \bm{\tilde{Z}^G} = \bm{Z^GM},  
\end{equation}
and the corresponding genetic random-effect vector becomes:
\begin{equation}
    \tilde{\bm{\alpha}} = \bm{M^{-1}\alpha}.
\end{equation}
We can check the covariance structure of $\tilde{\bm{\alpha}}$:
\begin{align}
cov(\tilde{\bm{\alpha}}) &= \bm{M}^{-1} (\bm{A} \otimes \bm{C^G}) (\bm{M}^{-1})^\intercal  \notag \\
&= (\bm{L}^{-1} \otimes \bm{I_K})(\bm{LL}^\intercal \otimes \bm{C^G})(\bm{L}^{-1} \otimes \bm{I_K})^\intercal \notag\\
&= (\bm{L}^\intercal \otimes \bm{C^G})(\bm{L}^{-1} \otimes \bm{I_K})^\intercal \notag \\
&= \bm{I_N \otimes C^G}
\end{align}
which is in a block diagonal structure. Through this parameterisation, we guarantee correlations between levels of genetic random effect while all subjects remain independent.

Now let us add the environmental random effect and rewrite Equation~({\ref{eqn:GFMEM3}}) in a more compact notation:
\begin{equation}\label{eqn:GMEM4}
    \bm{\tilde{y}} = \bm{X\beta + Z^*u^* + \epsilon}
\end{equation}
with the design matrix $\bm{Z^*} = \bm{[Z^GM}, \bm{Z^E]}$ and the random effects $\bm{u^*} = \bm{[M^{-1}\alpha}, \bm{\gamma]}^\intercal$. This model can now be fitted using standard implementation for mixed-effects models, such as \textbf{lme4} in R.

%% file: 3_JointFMEM.tex
\section{Joint model for amplitude and warping functions}\label{Sec:Joint}
In the previous section, we have described a model that can be used to analyse curves after alignment, i.e. to decompose additive genetic and environmental variability for the amplitude components of the FV traits. However, this approach fails to capture the association between amplitude and phase variations, and it doesn't allow us to explore the additive genetic and environmental components of the phase variability. One way to extend the model is to incorporate warping functions in mixed-effects models. \citet{lee_2016_combined} proposed the functional combined principal component analysis (functional combined PCA), which simultaneously captures the amplitude, phase and their association. One extension of the previous analysis is to use the combined FPCs as the basis to fit the genetic FMEM. Let us revisit the functional combined PCA from  \citet{lee_2016_combined} and discuss our modifications to analyse genetic data.

\subsection{Transformation for warping functions}
We assume a smooth random function $f$ consists of the amplitude function $y$ and the warping function $h$:
$$f(t) = y(h^{-1}(t)), \quad t \in [0,1].$$
As before, $y$ is assumed to be in the $L^2$ space, and $h$ is a strictly monotone function. The immediate problem is that $h$ is a constrained function and is not naturally representable using basis functions. Therefore, the first step is to transform the warping function into an unconstrained form in $L^2[0,1]$.
\citet{happ_2019_a} studied the joint amplitude and phase variations framework via FPCA. They presented several different transform approaches for warping functions and discovered that the centred log-ratio (CLR) transformation is preferred in the context of FPCA. The warping function can be interpreted as a cumulative distribution function of a continuous random variable. Taking the first derivative $h^\prime$ gives a unique probability density function on $[0,1]$. As shown in \citet{egozcue_2006_hilbert}, density functions, or more precisely, equivalence classes of density functions, form a vector space known as Bayes Hilbert space $B^2[0,1]$. We define the CLR transformation of $h^\prime$ as $\psi_{CLR}: B^2[0,1] \rightarrow L^2[0,1]$,
$$\psi_{CLR}(h^\prime(t)) = \log(h^\prime(t)) - \int_0^1 \log(h^\prime(t)) \,dt,$$
and the inverse transformation is 
$$h = \frac{\int_0^t \exp(\psi_{CLR}(s)) \,ds}{\int_0^1 \exp(\psi_{CLR}(s)) \,ds}.$$ 
We now refer to the transformed warping function as the phase function, and the combined functional PCA will be applied to both the amplitude and the unconstrained phase functions.

\subsection{Functional combined PCA}
To model the association between the amplitude and phase, \citet{lee_2016_combined} concatenated the amplitude function $y$ and the phase function $\psi$ using a scaling parameter $C$:
\begin{equation}\label{eqn:gc}
    g^C(t) = 
    \begin{cases} 
    y(t), & t \in [0,1), \\
    C\psi(t-1), & t \in [1,2].
    \end{cases}
\end{equation}
The parameter is introduced to adjust the scaling imbalance between $y$ and $\psi$. Assume $C$ is given, and denote the covariance function of $g^C$ by $\Sigma_{g^C}$, Mercer's theorem implies
\begin{equation}
     \Sigma_{g^C} = \sum_{k=1}^{\infty} \lambda_k \xi_k^C(t)\xi_k^C(s), \quad s,t \in [0,2],
\end{equation}
where $\lambda_k$ are the eigenvalues in descending order and $\xi_k^C$ are the corresponding orthogonal eigenfunctions. By Karhunen-Lo\'eve expansion,
\begin{equation}\label{eqn:cfpca}
    g^C(t) = \mu(t) + \sum_{k=1}^\infty S^C_{k}\xi_k^C(t),
\end{equation}
where $\xi^C_k(t)$ are FPCs and $S^C_{k} = \int(g^C(t) - \mu(t))\xi_k^C(t) \,dt$ are the FPC scores. The function $g^C$ is then divided into amplitude and phase parts as
\begin{gather}\label{eqn:cfpca2}
    \begin{aligned}
    y^C(t) &= \mu(t) + \sum_{k=1}^{\infty} S_k^C \xi_k^C(t), \quad t \in [0,1),\\
    \psi^C(t) &= \sum_{k=1}^{\infty} \frac{S_k^C}{C} \xi_k^C (t+1), \quad t \in [0,1].
    \end{aligned}
\end{gather}
The joint variation between $y$ and $\psi$ is paired in eigenfunctions $\xi_k^C$.

Here, we make two remarks on the scaling parameter $C$. First, $C$ is applied only to the phase part of $g^C$ to keep the original unit of the observed function $f$. Second, the eigenfunctions and eigenvalues vary for different values of $C$. For a small $C$, the first few eigenfunctions capture more variations from the amplitude variation. As $C$ increases, the leading eigenfunctions reflect more phase variations.

\subsubsection{Choice of \texorpdfstring{$C$}{C}}
\citet{lee_2016_combined} proposed to use $C$ to be dependent on the original function $f$ to succinctly represent the combined variations in the original function space. Let $K$ be a positive integer and let Equation~(\ref{eqn:cfpca2}) to be truncated to $K$-dimension:
\begin{gather}\label{eqn:cfpca3}
    \begin{aligned}
    y_K^C(t) &= \mu(t) + \sum_{k=1}^K S_k^C \xi_k^C(t), \quad t \in [0,1),\\
    \psi_K^C(t) &= \sum_{k=1}^K \frac{S_k^C}{C} \xi_k^C (t+1), \quad t \in [0,1].
    \end{aligned}
\end{gather}
For a given $K$, an optimal value of $C$ can be found by optimising the reconstructed function based on the first $K$ FPCs. A reconstruction of $f$ is given by $\hat{f}_K =  y_K^C \circ {h_K^C}^{-1}$, where $h_K^C = {\psi_K^C}^{-1}$ and ${h_K^C}^{-1}$ is the inverse of the reconstructed warping function. Thus, an optimal choice of $C$ is 
\begin{equation}
    \hat{C} = \arg \min_{C > 0} \mathbb{E}[\|\hat{f}_K - f \|_2].
\end{equation}
Several ad hoc procedures, such as the fraction of variance explained by the first few FPCs, can be used to determine the choice of $K$.

When analysing genetic data, we are more interested in the covariance functions than in reconstructing the original function from the data. Thus, we instead choose $C$ so that the amplitude and scaled phase functions reflect the same variability level. If we denote $\Sigma_y$ and $\Sigma^C_\psi$ as the amplitude and scaled phase covariance functions, we set $C$ such that they have the same norm.

\subsubsection{Estimation procedure}
Suppose that for each function $f_i$, the amplitude component $y_i(t)$ and phase component $\psi_i(t)$ are discretised on the time grid $0 = t_{i1} < \cdots < t_{ij} = 1$, i.e., $y_{ij} = y_i(t_{ij})$ and $\psi_{ij} = \psi_i(t_{ij})$. We compute the sample amplitude covariance function $\hat{\Sigma}_y$ and scaled phase covariance $\hat{\Sigma}^C_\psi$ on this time grid. Then, we compute $\hat{C}$ such that 
\begin{equation}\label{eqn:computeC}
\frac{\norm{\hat{\Sigma}_y}}{\norm{\hat{\Sigma}^C_\psi}} = 1, 
\end{equation}
and $\norm{\cdot}$ represents a matrix norm. 

For a given $\hat{C}$, the joint curve 
\begin{equation}
    \bm{g_i}^{\hat{C}} = 
\begin{bmatrix}
    \bm{y_i}\\
    \hat{C}\bm{\psi_i}
\end{bmatrix},
\quad
\bm{y_i} = [y_{i1}, \cdots, y_{ij}]^\intercal, \quad \bm{\psi_i} = [\psi_{i1}, \cdots, \psi_{ij}]^\intercal.
\end{equation}
The eigen-decomposition of the sample covariance function $\hat{\Sigma}_{g^{\hat{C}}}$ is
\begin{equation}
   \hat{\Sigma}_{g^{\hat{C}}} = \sum_{k=1}^{K} \hat{\lambda}_k^{\hat{C}} \hat{\bm{\xi}}_k^{\hat{C}} ( \hat{\bm{\xi}}_k^{\hat{C}} )^\intercal, 
\end{equation}
and $\hat{\bm{\xi}}_k^{\hat{C}}$ is an eigenvector of values $\hat{\xi}_k^{\hat{C}}(t_{ij})$. Estimates of the eigenfunctions $\hat{\xi}_k^{\hat{C}}$ are obtained using any convenient interpolation method. 

\subsection{Joint FMEM}\label{Subsec:JointFMEM}
Using the joint curves $g_i^{\hat{C}}$ as the response (trait) vector and the combined FPCs (eigenfunctions) $\hat{\xi}_k^{\hat{C}}$ as the functional basis, the joint FMEM follows the same structure as Equation~(\ref{eqn:GFMEM1}): 
\begin{equation}\label{eqn:JFMEM}
    g^{\hat{C}}_{ij} = \sum_{k = 1}^{K_f} \beta_k\hat{\xi}_k^{\hat{C}}(t_{ij}) + \sum_{k=1}^{K_g} \alpha_{ik}\hat{\xi}_k^{\hat{C}}(t_{ij}) + \sum_{k=1}^{K_e} \gamma_{ik}\hat{\xi}_k^{\hat{C}}(t_{ij}) + \epsilon_{ij},
\end{equation}
where $\beta_{ik}$ are the fixed-effects coefficients; $\alpha_{ik}$ and $\gamma_{ik}$ are the additive genetic and environmental random effects; $\epsilon_{ij}$ are the residual errors. Compared with the previous model, the combined eigenfunctions $\hat{\xi}_k^{\hat{C}}$ contain the association of amplitude and phase variations in genetic data, which provides insights into how curve registration affects the phenotypic variation instead of overlooking temporal dynamics. 

We want to make two remarks on the joint FMEM. First, the mathematics of the joint FMEM is the same as the previous case for aligned curves, so the joint model estimation follows the same procedure explained in Section~\ref{Subsec:estalign}. Second, when the functional data are observed on varying periods, a necessary step is to standardise time to $[0,1]$ with respect to the original (individual) interval $\mathcal{T}_i$. Therefore, we may need to include the interval $\mathcal{T}_i$ in the joint model to account for individual variability in time scaling. The joint curve, after discretisation, can be redefined as:
\begin{equation}\label{eqn:gc2}
    \bm{g_i}^{C} = 
    \begin{bmatrix}
    \bm{y_i}\\
    C\bm{\psi_i} \\
    \mathcal{T}_i
    \end{bmatrix}
\end{equation}
Equation~(\ref{eqn:gc2}) preserves the amplitude and phase information while adding a scalar (time interval) in a multivariate setting. The extended eigenfunctions (FPCs), obtained from a multivariate PCA, are
\begin{equation}
    \bm{\tilde{\xi}_k^C} = 
     \begin{bmatrix} 
        \bm{\xi_k^A} \\
        \bm{\xi_k^P} \\
        \xi^{\mathcal{T}}_k
    \end{bmatrix}.
\end{equation}
The combined FPCs present a joint pattern in which amplitude shape, phase variation and the curve duration together, with the final component measuring variation, such as the overall time-scaling effect, that is not explicitly represented in the curves.

%% file: 4_Simulation.tex
\section{Simulation study}\label{Sec:Simulation}
In this section, we investigate the performance of the estimation procedure for the genetic functional mixed-effects models on simulated data. The simulated scenarios have been chosen to mimic the red flour beetle data (see more details in Section~\ref{Sec:TCData}). Since the way the joint model parameters are estimated follows the same methodology as the model for aligned curves, only with higher dimensional data, i.e., joint curves of amplitude and phase functions, we conduct simulation studies using the aligned model, which is computationally less intensive. 

\subsection{Simulation settings}
Consider the synthetic data generated from Model~(\ref{eqn:GFMEM3}). We use five cubic B-spline functions as the model basis and, for simplicity, set equal covariance for both additive genetic and environmental effects. 
Let $\bm{\phi(t)}$ be basis function. The additive genetic and environmental covariance functions which will be estimated are: 
\begin{align*}
G(t,s) &= \phi(t)^{\intercal} \bm{C^G} \phi(s) \\
E(t,s) &= \phi(t)^{\intercal} \bm{C^E} \phi(s).
\end{align*}

\begin{figure}
     \centering
     \begin{subfigure}[h]{0.495\textwidth}
         \centering
         \includegraphics[width=\textwidth]{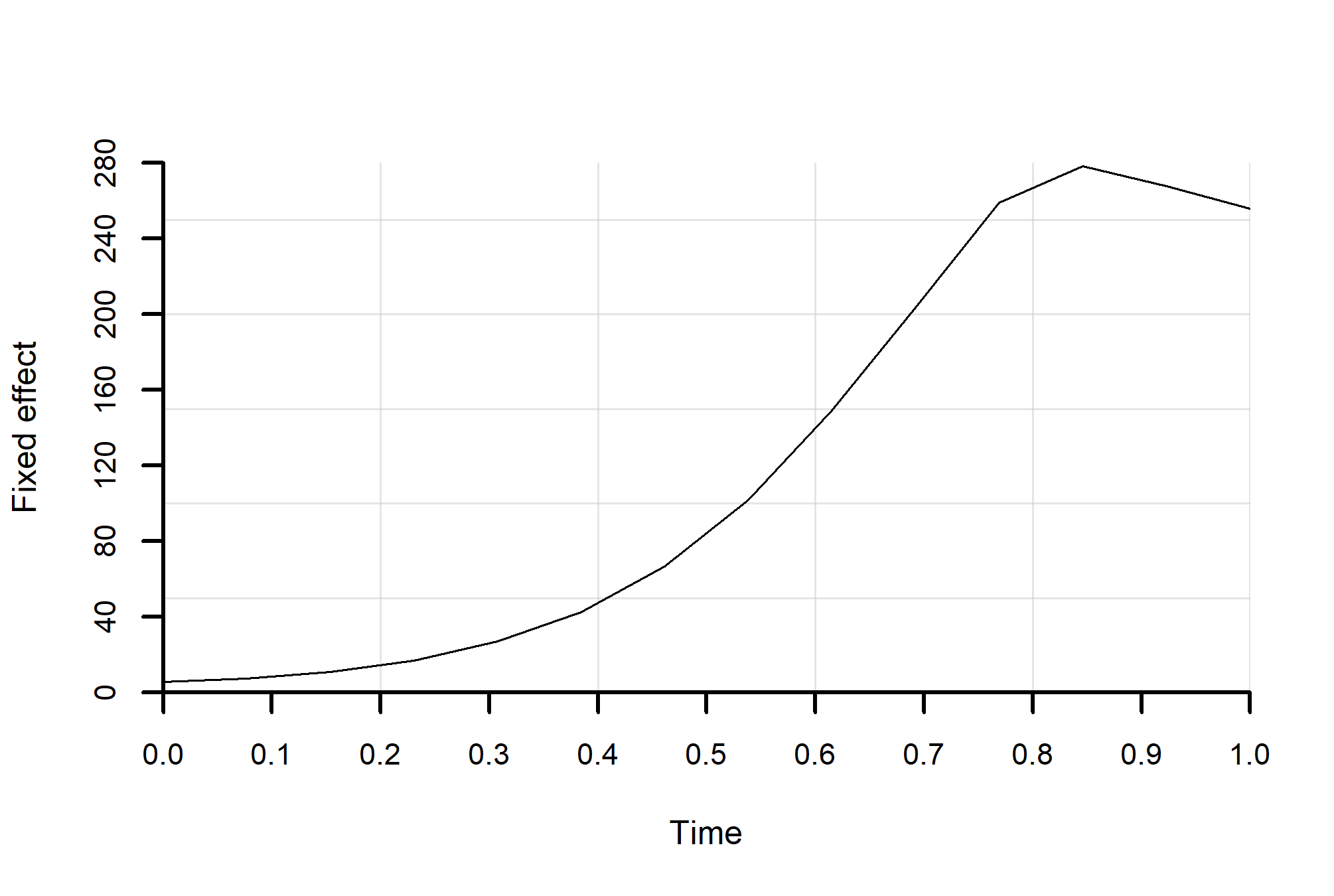}
         \caption{}
         \label{fig:TrueFE}
     \end{subfigure}
     \begin{subfigure}[h]{0.495\textwidth}
         \centering
         \includegraphics[width=\textwidth]{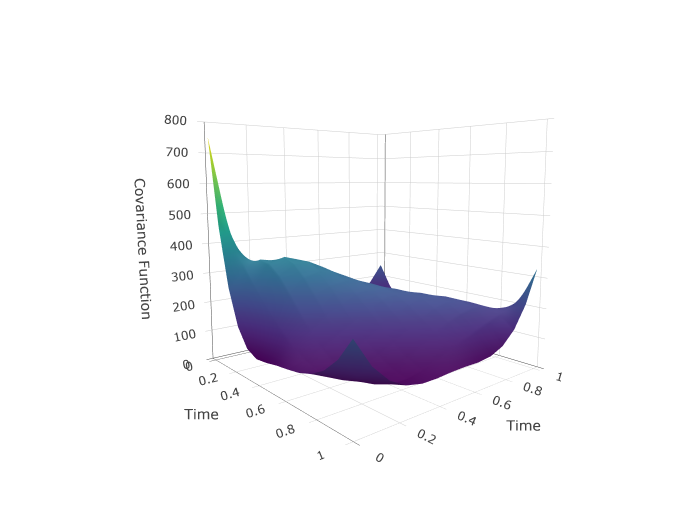}
         \caption{}
         \label{fig:truecov}
     \end{subfigure}
     \caption{ Parameters of the simulation study. (a) True fixed effect. (b) Surface plot of the true covariance functions $G(t,s)$ and $E(t,s)$.}
     \label{fig:sim}
\end{figure}

The random effect parameters $\bm{\alpha}$ and $\bm{\gamma}$ are assumed to follow a multivariate normal distribution with covariance $\bf{A} \otimes \bf{C^G}$ and $\bf{I_N} \otimes\bm{C^E}$, respectively. In our simulation study, we build a dataset with 873 individuals, so the same relationship matrix $\bf{A}$ computed from the \textit{Tribolium castaneum} can be used in our simulations. This is to use a realistic relationship structure that is used in full-sib/half-sib genetic breeding experiments. Each curve is simulated with 14 regular observations within the unit interval. The fixed effect is set to the sample mean of the \textit{Tribolium castaneum} data. Additionally, the residual variance $\sigma^2_{\text{res}}$ is varied and set to 15. Figure~\ref{fig:sim} shows the plot of the true fixed effect and covariance function. 

We generate 50 datasets and fit simulated data to our genetic FMEM with FPCs as basis functions. We use 6 FPCs to estimate the fixed effect and 5 FPCs for the random effects. FPCs capture the directions of the largest variability in the data, emphasising specific variation directions. However, the fixed effect reflects the population mean, representing the overall trend in the data. We found that in simulations, a few additional FPCs help capture the broader general trend more accurately than random effects.  

We compare the estimated fixed effects, covariance functions and the genetic eigenfunctions to the true ones, with the uncertainty of functional estimators assessed using simultaneous bootstrap confidence bands \citep[see][]{cuevas_2006_on}. 

\subsection{Simulation results}
We begin by visualising the estimated functions using a functional boxplot, which displays the functional median, central (50\%) envelope, and outliers based on the centre-outward ordering induced by (modified) band depth for functional data. This concept is well-explained in \citet{lopezpintado_2009_on} and \citet{sun_2011_functional}.

\begin{figure}[h]
     \centering
     \begin{subfigure}{0.32\textwidth}
         \centering
         \includegraphics[width=\textwidth]{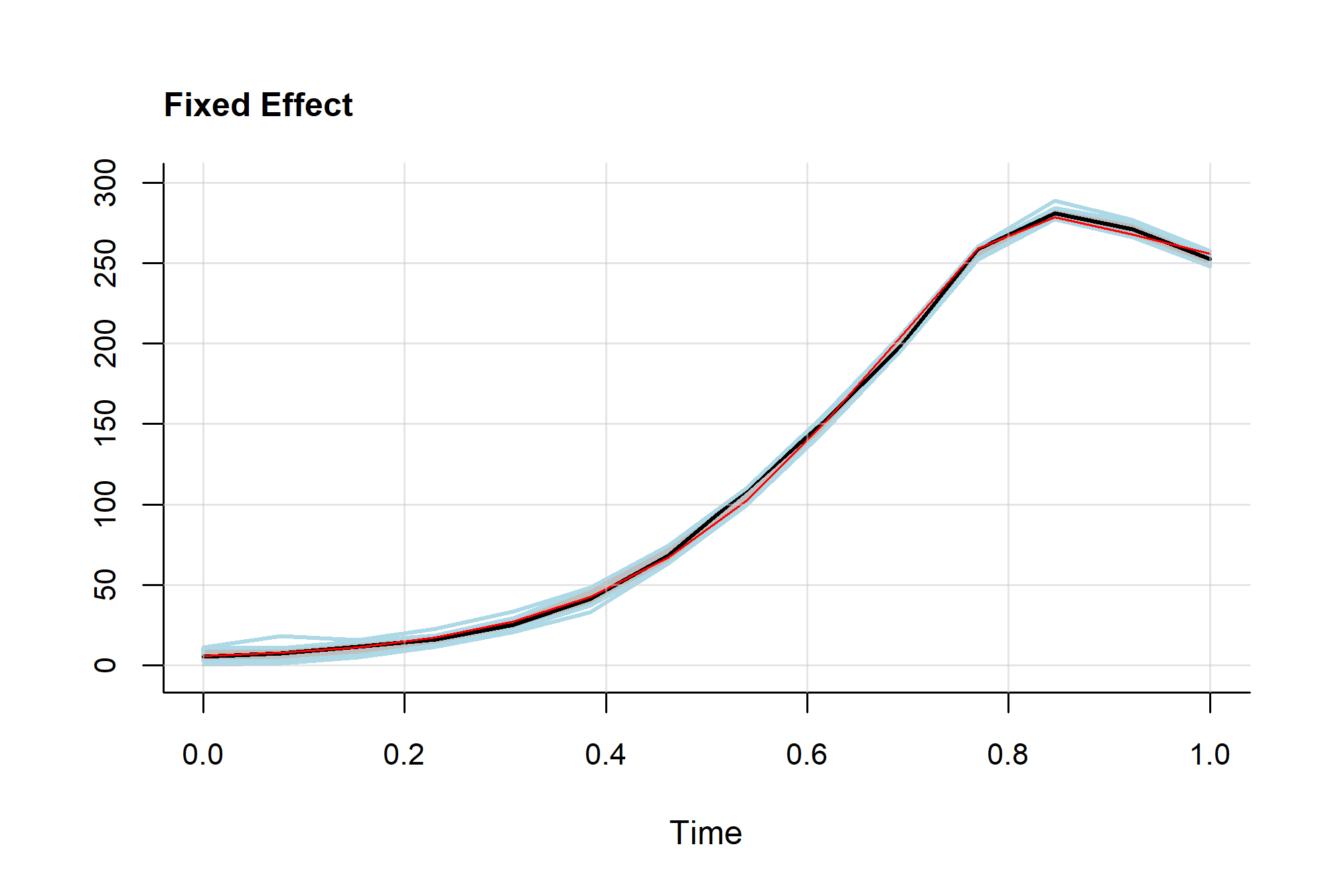}
         \caption{}
         \label{fig:FEfunboxplot}
     \end{subfigure}
     \begin{subfigure}{0.32\textwidth}
         \centering
         \includegraphics[width=\textwidth]{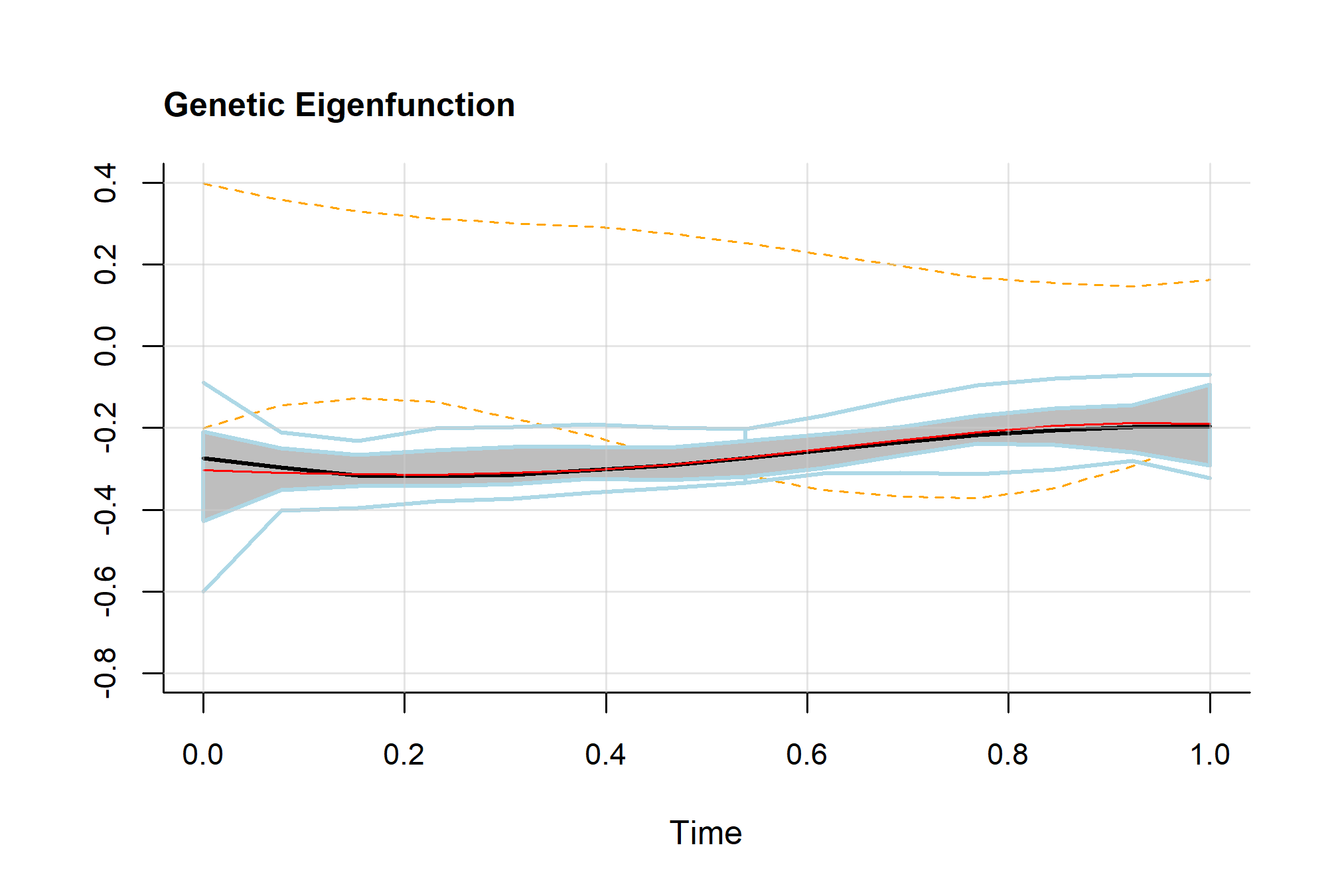}
         \caption{}
         \label{fig:FunGebox14}
     \end{subfigure}
     \begin{subfigure}{0.32\textwidth}
         \centering
         \includegraphics[width=\textwidth]{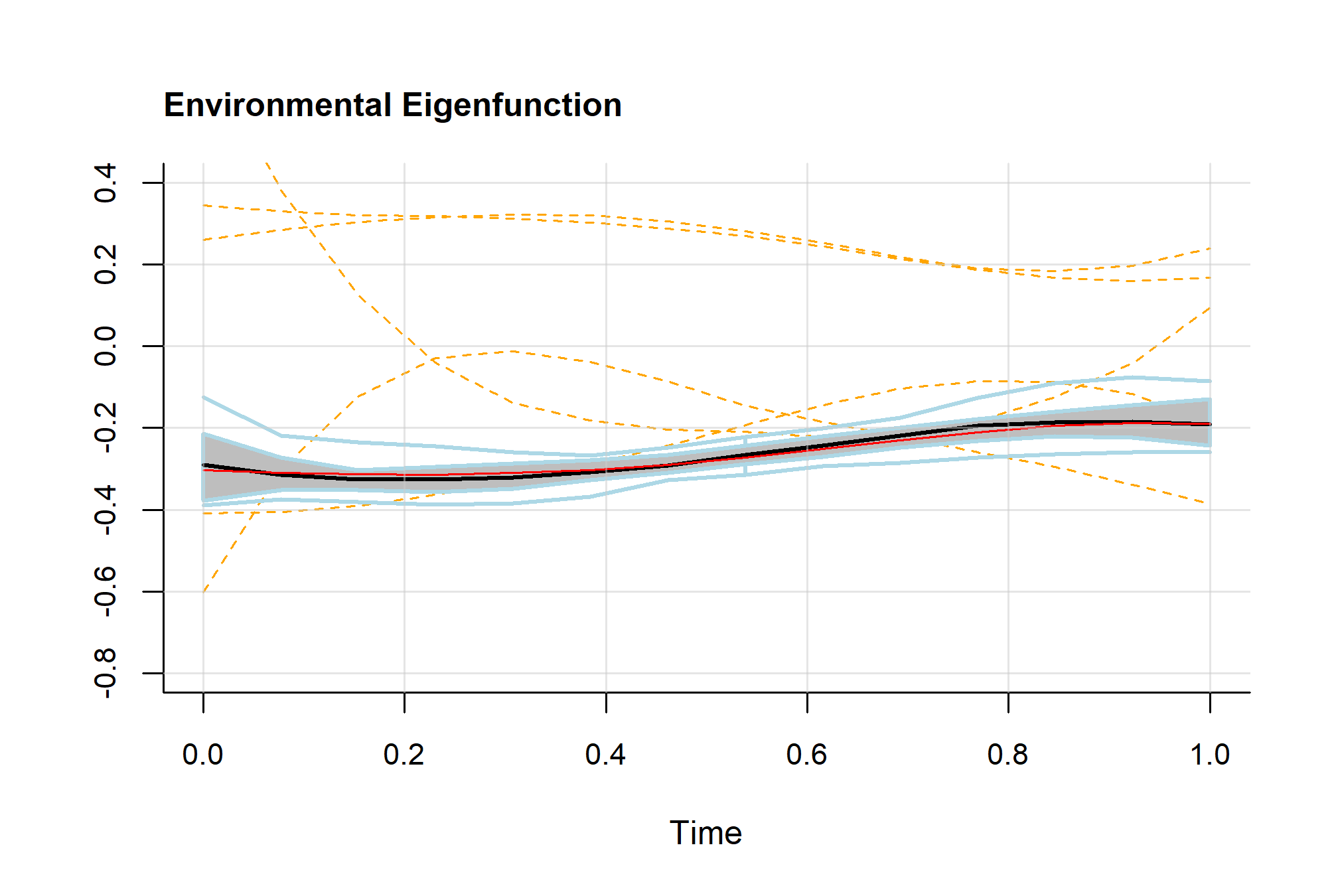}
         \caption{}
         \label{fig:FunEebox14}
     \end{subfigure}
     \caption{Functional boxplots for the (a) fixed effect, (b) additive genetic eigenfunctions, and (c) environmental eigenfunction based on 50 samples. Each panel displays the true function (red solid line), the median of the estimated samples (black solid line), and the 50\% central region (light grey shaded band). The dashed orange lines are outliers.} 
     \label{fig:eigenfunboxplot}
\end{figure}

\begin{itemize}[align=parleft,left=0pt..1em]
    \item Fixed effect estimation

    Figure~\ref{fig:FEfunboxplot} displays functional boxplots of the estimated sample fixed effect. The red line in each subplot represents the true fixed effect, while the black line denotes the median curve of the sample fixed effects. The 50\% central region is shown in light grey. The median curve closely matches the true value, suggesting that our model provides stable and accurate estimates for the fixed effect.

    \item{Eigenfunction estimation}

    Figure~\ref{fig:FunGebox14} and Figure~\ref{fig:FunEebox14} show functional boxplots of the first eigenfunctions of the additive genetic and environmental covariance functions. Comparing the true eigenfunctions (red lines) with the median curves of the sample eigenfunctions (black lines), we see that the model reasonably estimates both additive genetic and environmental eigenfunctions. However, the 50\% central regions for the additive genetic eigenfunctions are slightly wider than those for the environmental ones, indicating greater variability in estimating the genetic covariance function. The orange dashed lines are outliers primarily caused by singular-fitted models, i.e., the estimated variance-covariance matrices are less than full rank \citep{bates_2015_fitting}. This issue leads to numerical instability, usually caused by overfitting with redundant basis functions. When singularity occurs, one solution is to refit the model with fewer basis functions. 
\end{itemize}

We then adapt the procedure for constructing bootstrap simultaneous confidence bands to visualise the distribution of additive genetic and environmental covariance functions, similar to a functional boxplot. By doing so, we capture the overall variability of functional outcomes across all simulations. We refer to this approach as the functional distribution band, distinguishing it from formal confidence bands used for inferential purposes.

\begin{figure}[h]
     \centering
     \includegraphics[width = \textwidth]{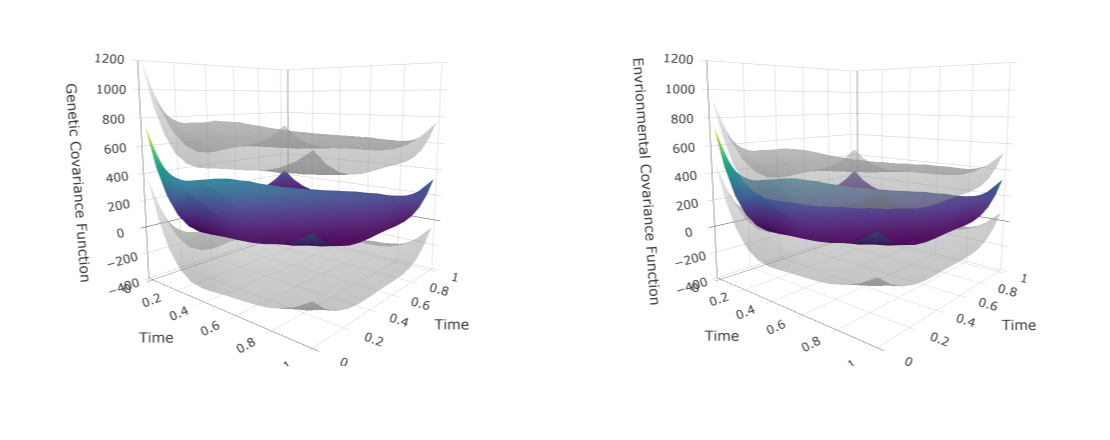}
     \caption{95\% functional distribution bands of the genetic (left) and environmental (right) covariance functions based on 50 samples. The blue-purple surface in each plot represents the true covariance function, and the grey surfaces are 95\% quantile boundaries.}
     \label{fig:RESCB}
\end{figure}

As shown in Figure~\ref{fig:RESCB}, the true additive genetic and environmental covariance functions (blue-purple surfaces) lie within the distribution bands (grey surfaces), although a wider band for the genetic function is observed.

\subsection{Bootstrap simulation results}
We use a parametric bootstrap procedure to evaluate the uncertainty of model estimates and construct simultaneous confidence bands for all groups of simulated datasets. As explained in the previous section, singular fitted results exist because of numerical instability. We exclude such results in the bootstrap analysis. A new dataset is generated to replace the discarded one until we have 50 groups of non-singular estimations, which guarantees the simulated datasets used in the bootstrap study have well-defined covariance structures. For each of the non-singular fitted models, 300 bootstrap samples are obtained after removing singular replicates and outliers. The 95\% simultaneous confidence bands for the fixed effect, additive genetic covariance function, environmental covariance function and their first eigenfunctions are constructed. Table~\ref{coverate} shows the empirical bootstrap coverage percentage for all functional estimators. The selected bootstrap replicates generate a narrow confidence band for the fixed effect, resulting in a coverage rate slightly lower than 95\%. The empirical coverage rates for eigenfunctions and covariance functions exceed the 95\% level. This may be due to having a bootstrap sample too small, but it is important to remember that the bootstrap bands are designed to cover the full surface, so it is not surprising that some specific directions have higher than nominal coverage.

\begin{table}[h!]
\centering
\begin{tabular}{>{\raggedright\arraybackslash}p{2.5cm} >{\raggedright\arraybackslash}p{3cm} >{\raggedright\arraybackslash}p{3cm} >{\raggedright\arraybackslash}p{2.5cm} >{\raggedright\arraybackslash}p{3cm}}
\toprule
\textbf{Fixed effect} & \textbf{Genetic eigenfunction} & \textbf{Environmental eigenfunction} & \textbf{Additive genetic covariance function} & \textbf{Environmental covariance function} \\
\midrule
0.92 & 0.98 & 1 & 1 & 0.96 \\

\bottomrule
\end{tabular}
\caption{Coverage percentage of the simultaneous confidence bands.}
\label{coverate}
\end{table}

\begin{figure}[h]
     \centering
     \begin{subfigure}[h]{0.32\textwidth}
         \centering
         \includegraphics[width=\textwidth]{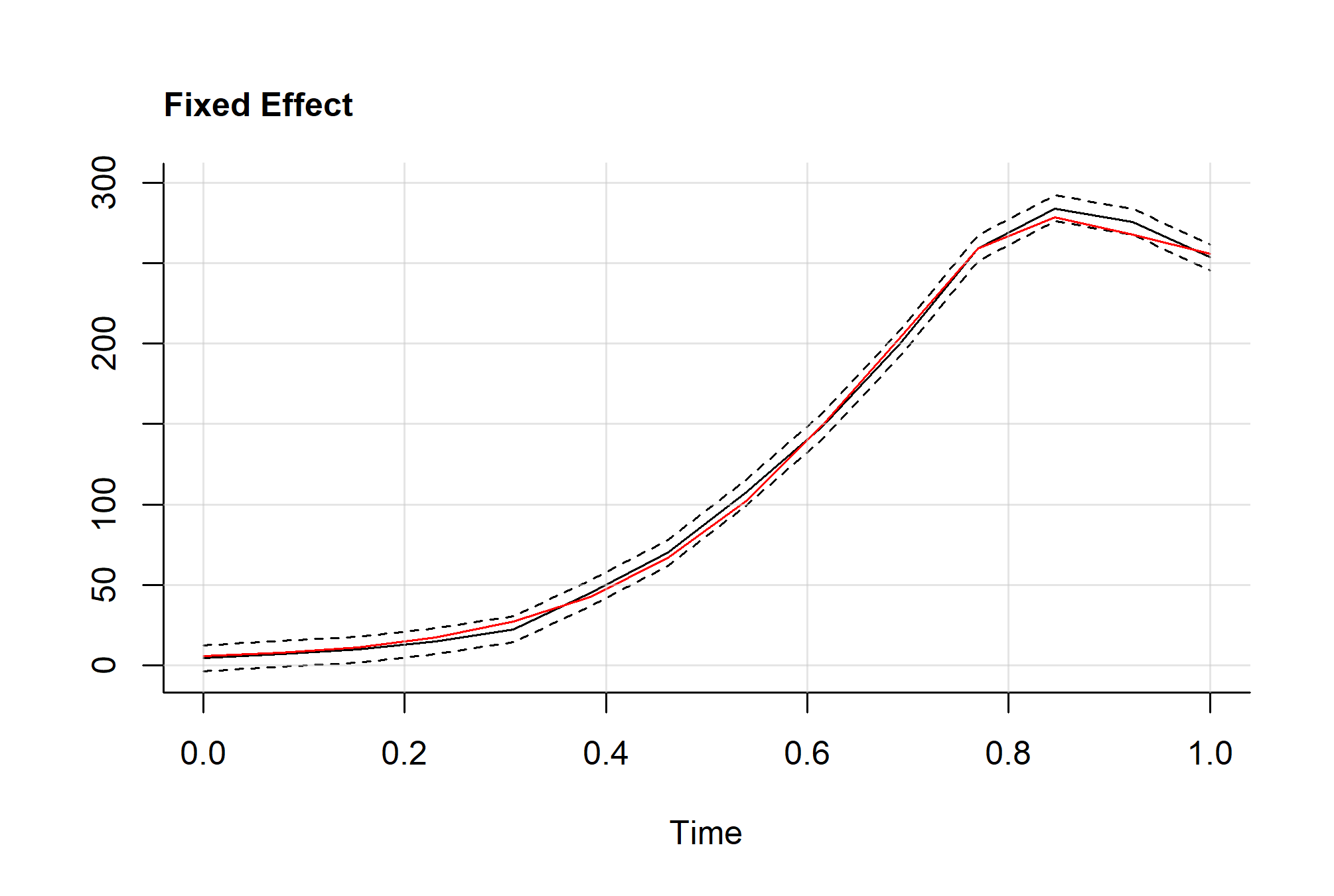}
         \caption{}
         \label{fig:bs41FESCB}
     \end{subfigure}
     \begin{subfigure}[h]{0.32\textwidth}
         \centering
         \includegraphics[width=\textwidth]{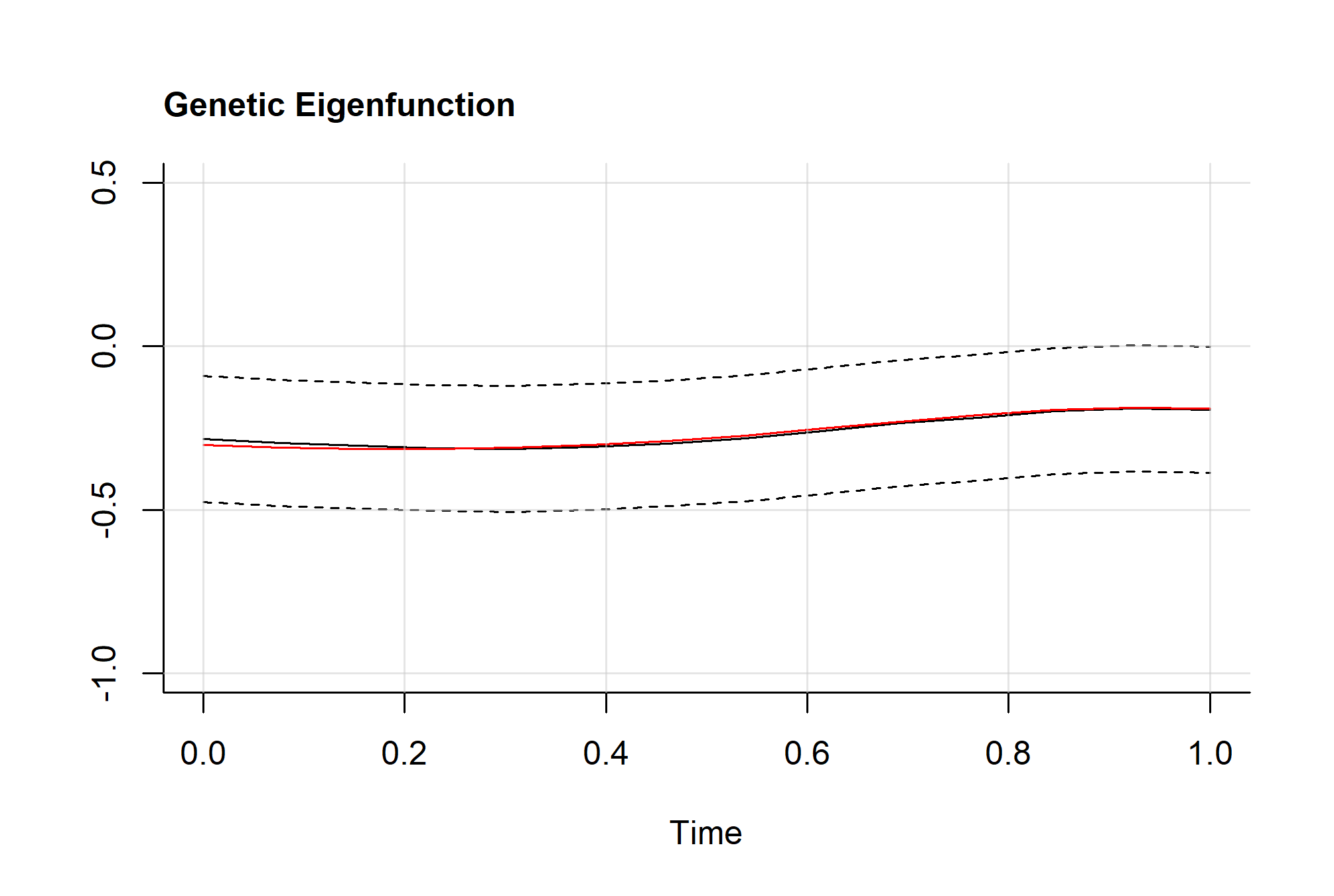}
         \caption{}
         \label{fig:bs41GESCB300}
     \end{subfigure}
     \begin{subfigure}[h]{0.32\textwidth}
         \centering
         \includegraphics[width=\textwidth]{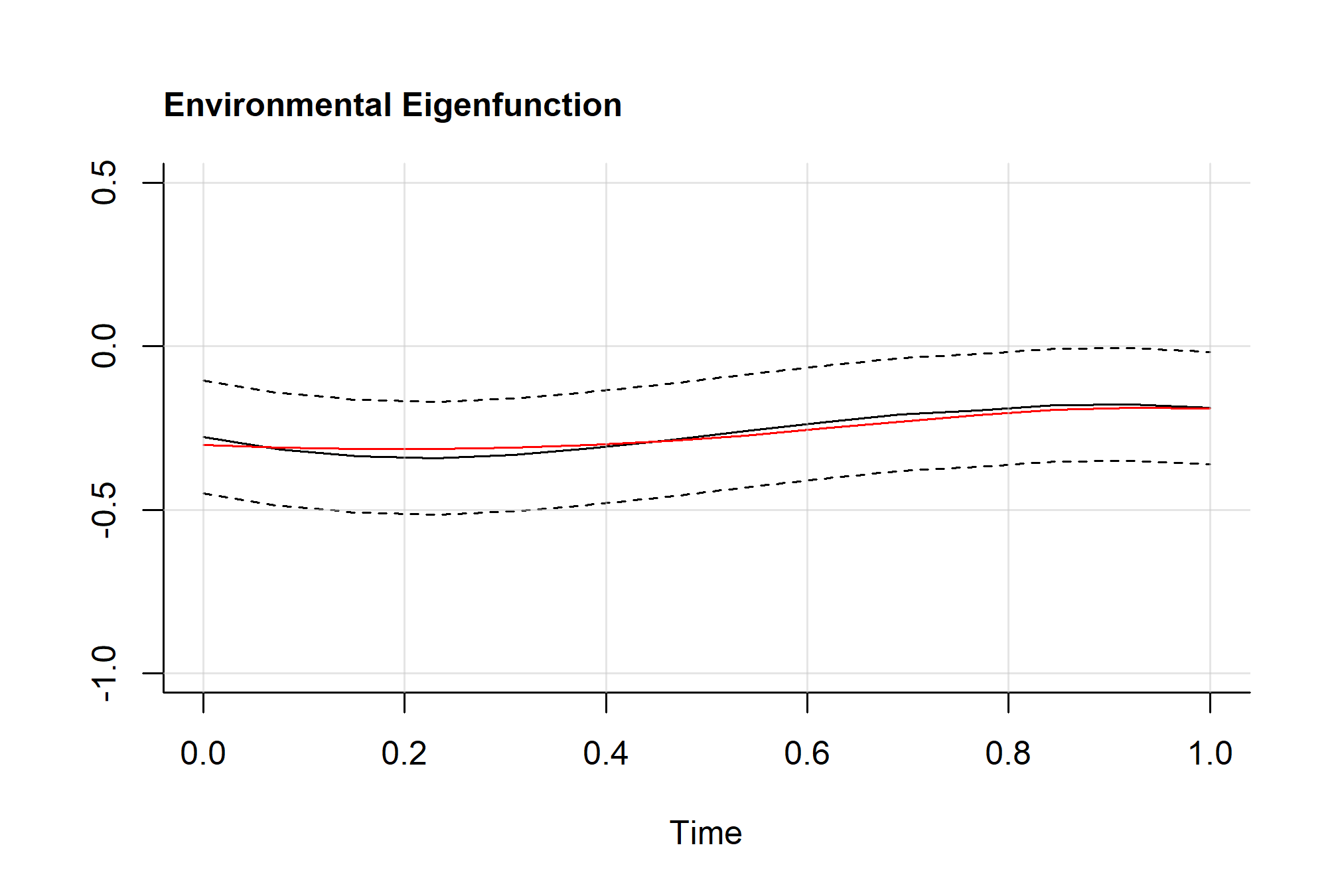}
         \caption{}
         \label{fig:bs41EESCB300}
     \end{subfigure}
     \caption{95\% simultaneous confidence bands of the (a) fixed effects, (b) additive genetic eigenfunction and (c) environmental eigenfunction based on 300 bootstrap samples. Each panel displays the true function (red solid line) and the estimated result from the fitted model (black solid line). The black dashed lines are 95\% quantile boundaries.}
     \label{fig:bsFE300}
\end{figure}

\begin{figure}[h]
     \centering
     \includegraphics[width = \textwidth]{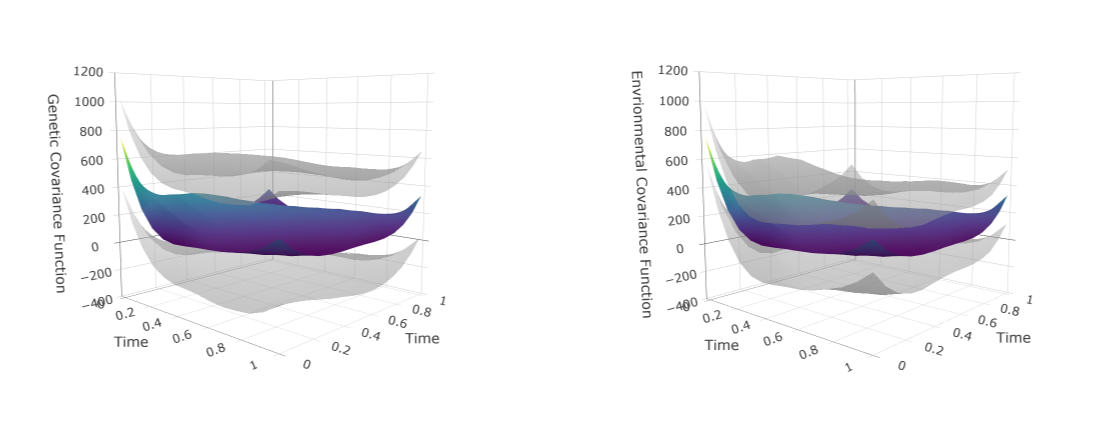}
     \caption{95\% simultaneous confidence bands of the additive genetic (left) and environmental (right) covariance functions based on 300 bootstrap samples. The blue-purple surface represents the true covariance function, and the grey surfaces are 95\% quantile boundaries.}
     \label{fig:bsRESCB}
\end{figure}

We present a bootstrap simulation for one simulated dataset (ID 41), which gives the best estimation of the additive genetic covariance function. The 95\% simultaneous confidence bands for bootstrap fixed effects are shown in Figure~\ref{fig:bs41FESCB}. The result indicates that our model effectively estimates the fixed effect (which is closely matched with the true value), though there is a slight deviation at the end of the interval. Figure~\ref{fig:bs41GESCB300},  Figure~\ref{fig:bs41EESCB300} and Figure~\ref{fig:bsRESCB} show that the true additive genetic and environment covariance functions and their first eigenfunctions lie in the 95\% confidence bands. 

%% file: 5_RealData.tex
\section{Real data analysis}\label{Sec:TCData}
In this section, we apply the aligned and joint models to the red flour beetle growth curves (\cite{irwin_2013_constraints}). 

\subsection{Red flour beetle growth curves}\label{Subsec:TCData} 
We consider the growth trajectories of the flour beetle $\textit{Tribolium castaneum}$ previously analysed by \citet{irwin_2013_constraints}. $\textit{Tribolium castaneum}$ is a common insect model for population genetic and development studies. Here, we briefly introduce the breeding design for the experiment that generated the data (see \cite{irwin_2013_constraints} for details).

\begin{figure}[h]
    \centering
    \begin{subfigure}{0.495\textwidth}
        \centering
        \includegraphics[width=\textwidth]{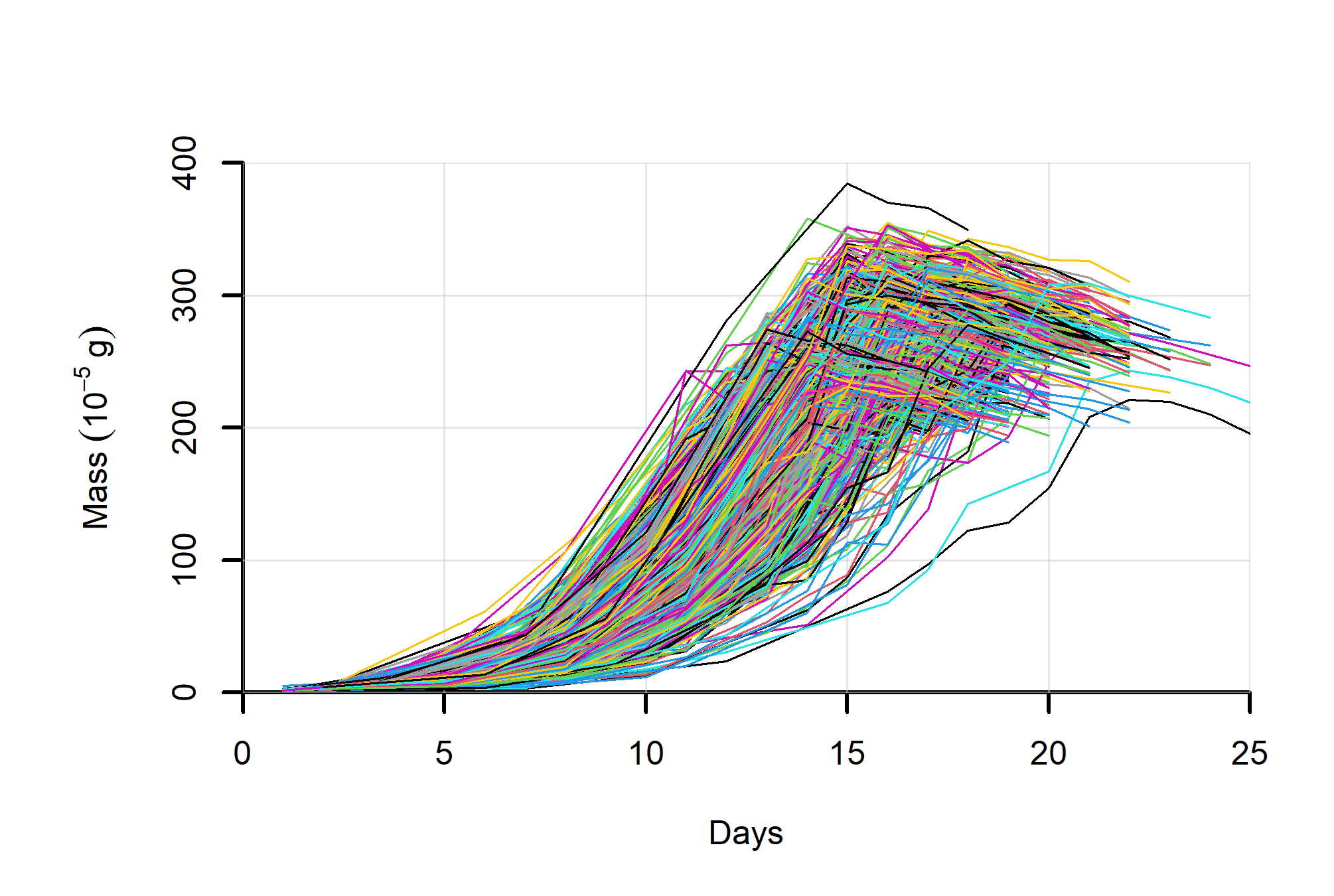}
        \caption{}
        \label{fig:rawgrowth}
    \end{subfigure}
    \begin{subfigure}{0.495\textwidth}
        \centering
        \includegraphics[width=\textwidth]{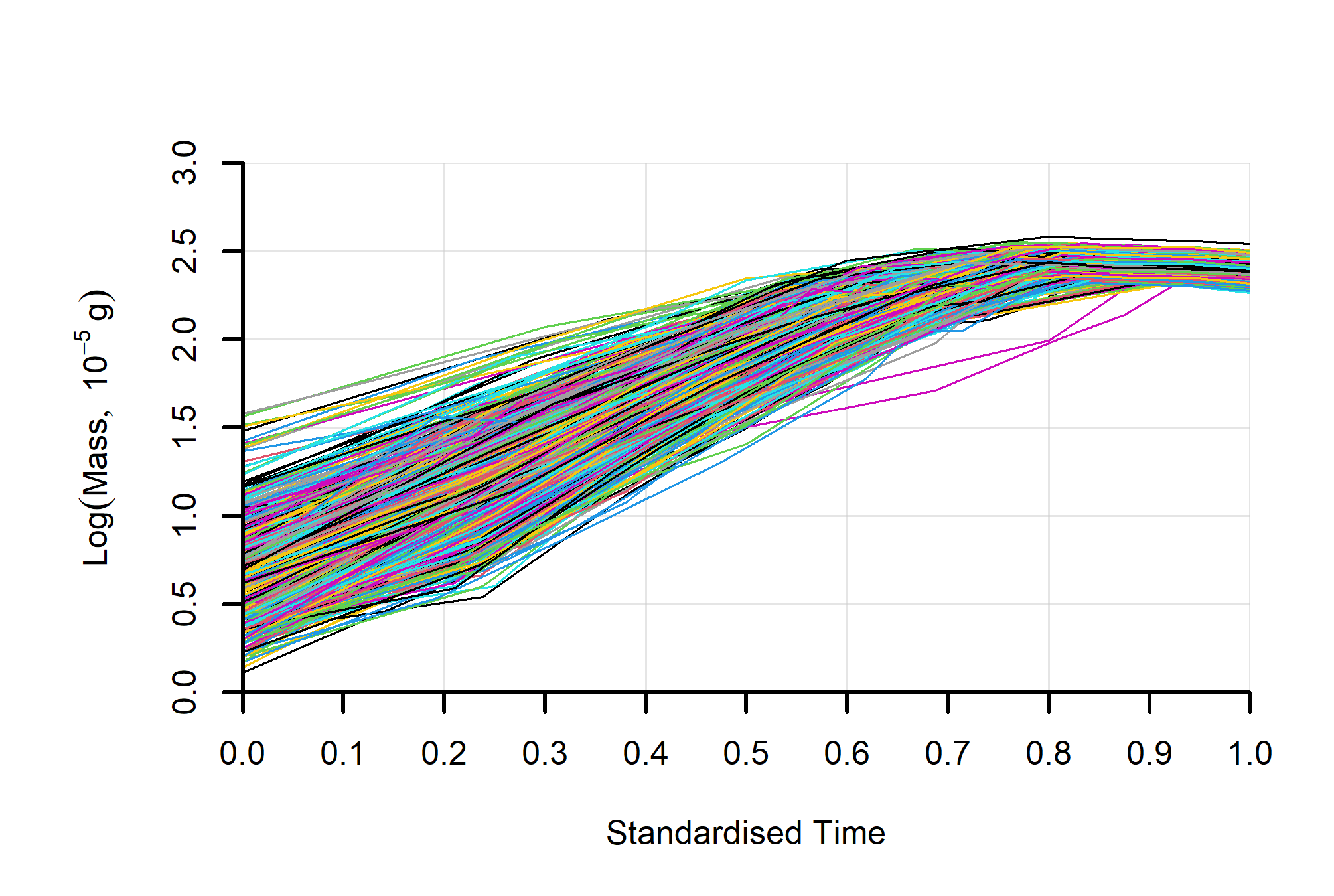}
        \caption{}
        \label{fig:lograwgrowth}
    \end{subfigure}
    \caption{Flour beetle growth curves. (a) Raw mass curves. (b) Log-mass curves rescaled to standardised time.}
    \label{fig:growth}
\end{figure}

All $\textit{Tribolium}$ individuals were derived from a stock population of the cSM++ strain, with approximately 300 individuals initially divided into 17 populations of about 20 individuals each and allowed to breed freely. Offspring were collected from these populations as they pupated, their sex was determined and recorded, and they were isolated into separate vials and used to create a stock of isolated, virgin adults. 

A half-sib/full-sib breeding design was used to facilitate quantitative genetic analysis. Full siblings share the same biological mother and father. Maternal half-siblings share the same mother only, and similarly, paternal half-siblings share the same father only \citep{Tanskanen2017}. From the stock of isolated virgin adults, one randomly chosen male was mated with five randomly chosen females, none of which were his siblings. This was repeated 30 times, using a total of 30 males and 150 females, thereby producing 30 half-sib families and 150 full-sib families. The sire, dam, and subjects' IDs were recorded and used to build the pedigree, which was then used to compute the relationship matrix $\bm{A}$. 

Body mass (in \SI{e-5}{\gram}) was measured for each larva at different ages during the larval period, and mass at pupation was included as the final mass measure for each growth curve. Individuals who did not reach pupation, those with $<$ 5 measures, or those who had pupated later than 22 days were excluded. Eventually, there were 873 qualified offspring with a total of 6860 measurements. The flour beetle growth curves are shown in Figure~\ref{fig:rawgrowth}. The unusual pattern of these curves is the decline in mass around day 15, as larvae reach their peak mass and then stop eating to begin searching for a place to pupate. This inflexion point provides a unique landmark, which can be used in curve alignment as a consistent reference across individuals.




\subsection{Data pre-processing}
Body mass in the \textit{Tribolium castaneum} dataset is not taken at a fixed frequency during the larval period. The largest number of measurements per individual is 14, and the smallest is 5. Mass is measured more frequently in the last days of the larval period. We first standardise time domains in proportion to the pupation time. Each growth curve is now measured from 0 (hatching) to 1 (pupation).   

We use a cubic smoothing spline as the basis function to transform discrete data into smooth functions. Smoothing parameters are selected by generalised cross-validation (GCV). However, the irregular sampling points lead to two problems if we directly smooth curves on the original scale: for subjects with fewer measurements, fitted curves approach those produced by standard linear regression; for subjects with sparse measurements around the starting period, the fitted curves have negative values near $t=0$. To address these problems, growth curves are $\log_{10}$-transformed before further analysis, which is a standard transformation in biological studies of size and growth. The log transformation imposes a positive smoothing constraint for fitted curves. It also improves numerical stability by reducing the large scaling effects in the dataset where body mass varies significantly between the early and later times. The smoothed curves are predicted onto a fine time grid in the unit interval.

\begin{figure}[h]
    \centering
    \begin{subfigure}{0.495\textwidth}
        \centering
        \includegraphics[width=\textwidth]{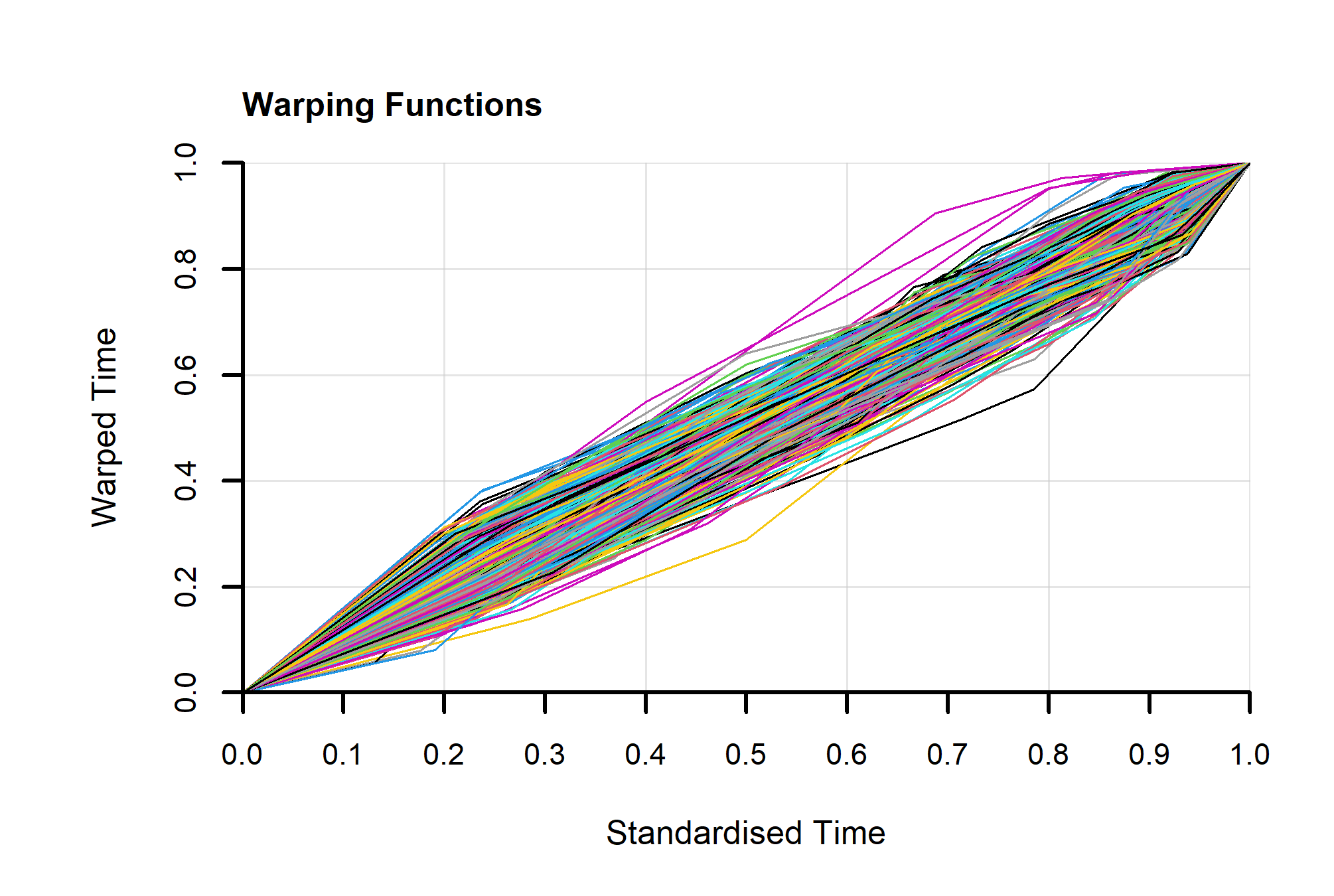}
    \end{subfigure}
    \begin{subfigure}{0.495\textwidth}
        \centering
        \includegraphics[width=\textwidth]{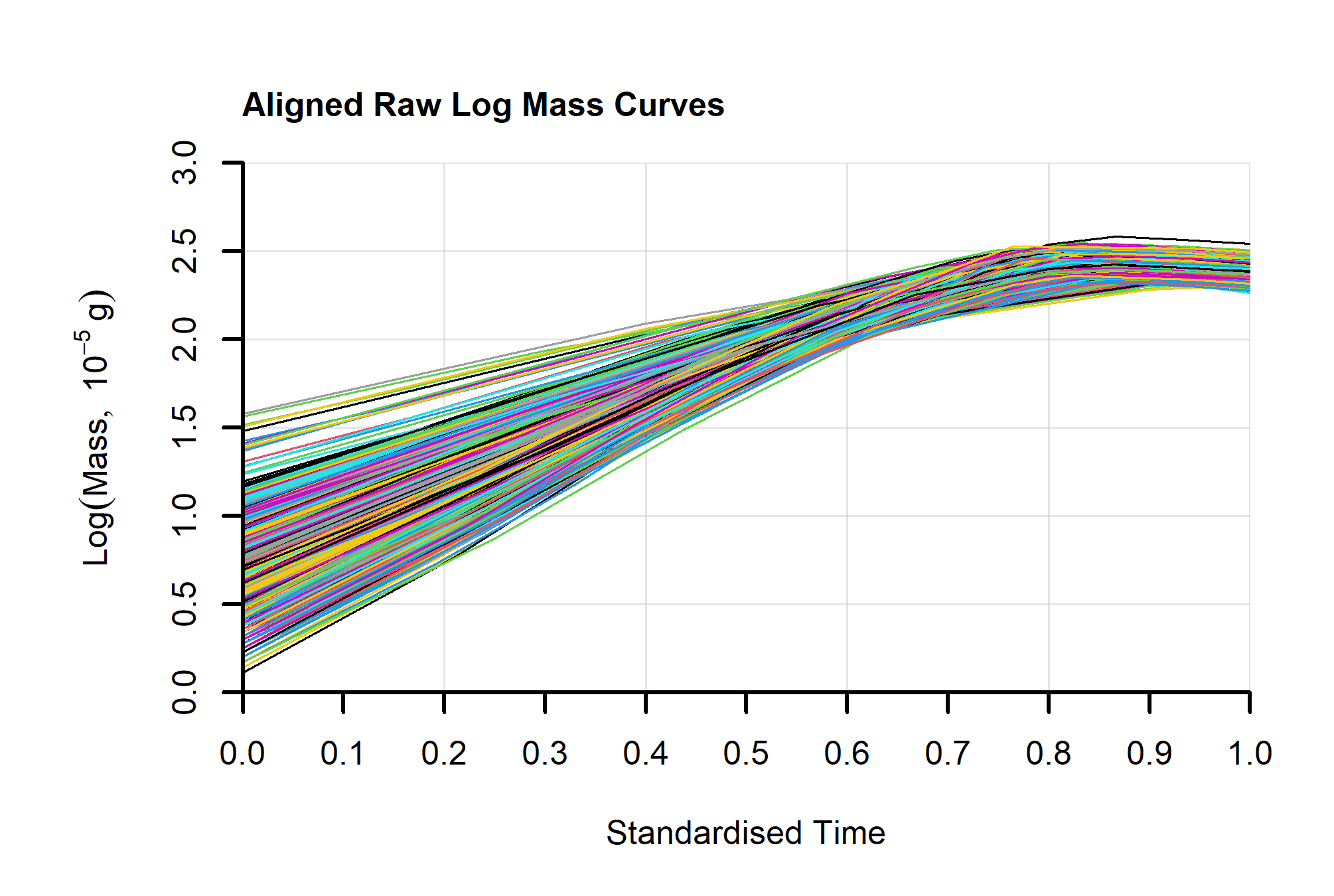}
    \end{subfigure}
    \caption{Align raw log-growth curves using warping functions.}
    \label{fig:warpinglog}
\end{figure}

As growth curves show amplitude and phase variations, we consider the continuous registration based on the Fisher-Rao metric proposed by \citet{srivastava_2011_registration} and implemented in R in the package \textbf{fdasrvf}. Model~(\ref{eqn:GFMEM1}) assumes random measurement errors are recorded in the realisation process. When smoothing the observed growth curves, essentially, these errors are smoothed out. Nonetheless, aligning the smoothed curves on a fine time grid could introduce collinearity, potentially making the model singular. To avoid this issue, we register the raw growth curves using warping functions interpolated to the original sampling points and then fit the aligned raw data to the genetic FMEM. The aligned raw growth curves on a logarithmic scale and the warping functions are displayed in Figure~\ref{fig:warpinglog}.

\begin{figure}[h]
    \centering
    \begin{subfigure}{0.495\textwidth}
        \centering
        \includegraphics[width=\textwidth]{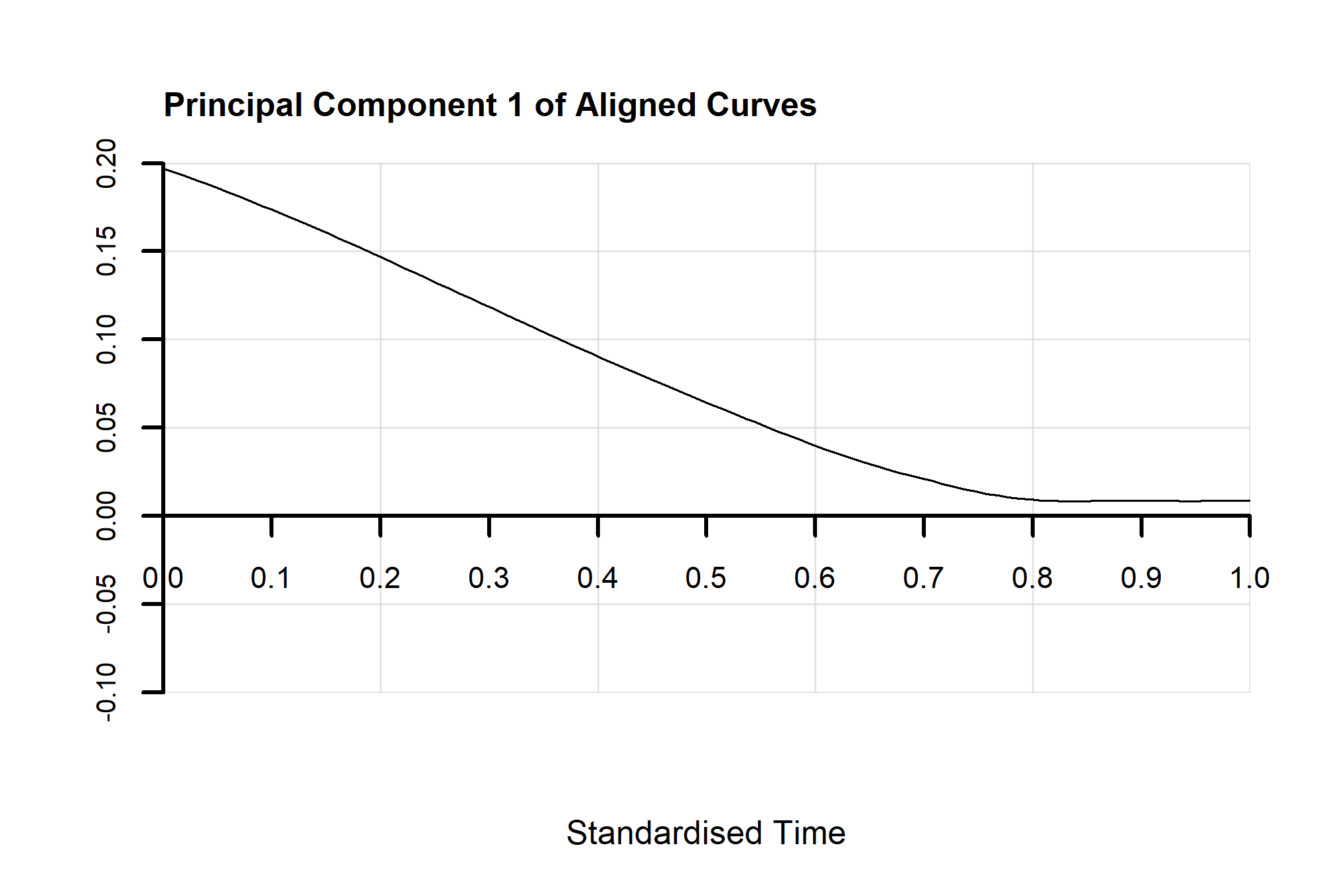}
    \end{subfigure}
    \begin{subfigure}{0.495\textwidth}
        \centering
        \includegraphics[width=\textwidth]{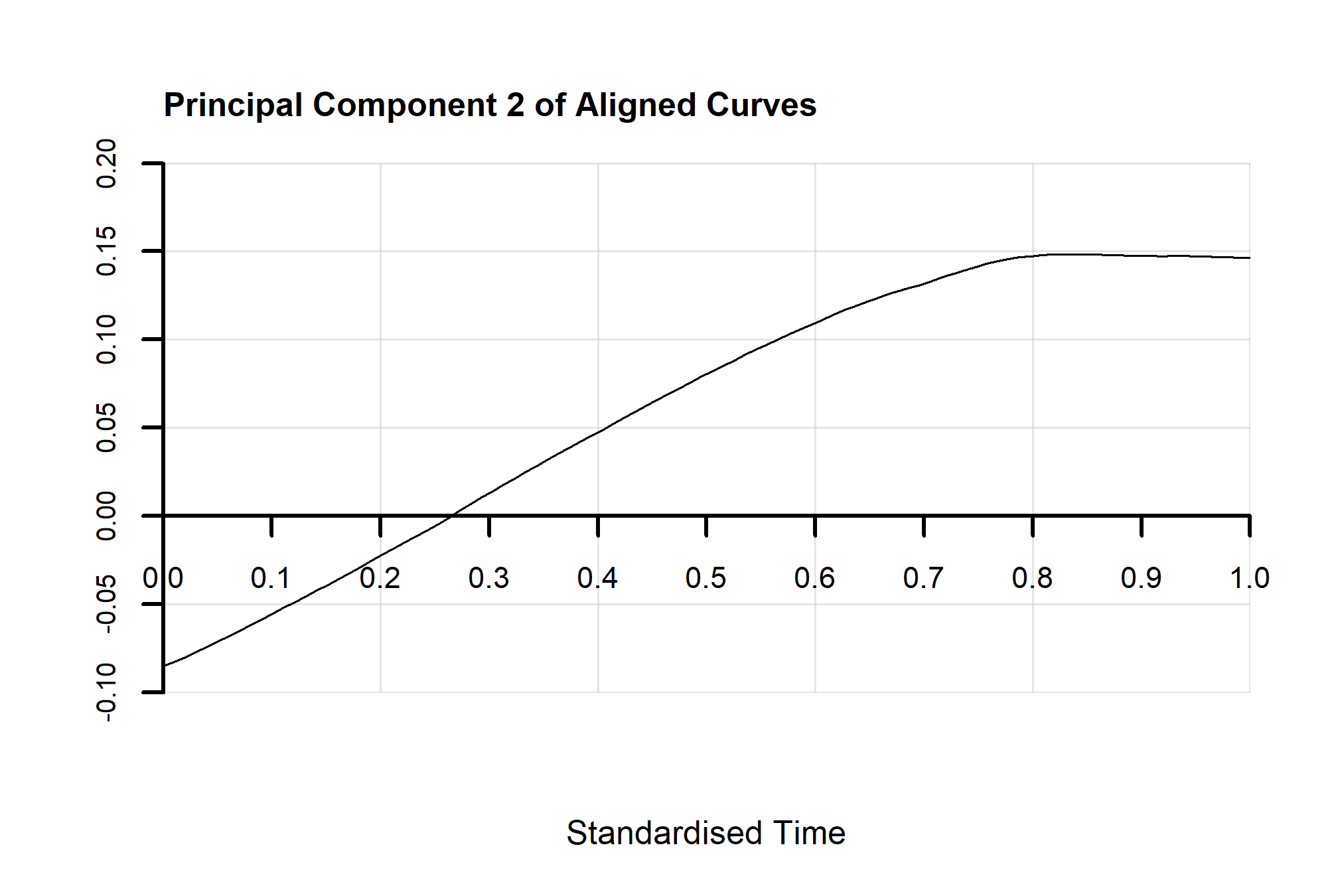}
    \end{subfigure}
    \caption{Plot of the first two FPCs of the smoothed aligned log-growth curves.}
    \label{fig:alignedFPClog}
\end{figure}

Then, we run FPCA on the smoothed and aligned curves. The first two FPCs (shown in Figure~\ref{fig:alignedFPClog}) capture over 99\% variance in the log-growth curves. Thus, we fit our genetic FMEM using two FPCs for both additive genetic and environmental components. Similarly, in the simulations, we use three FPCs (one more than the random effects) to present the mean shape. All FPCs computed from the smoothed-aligned curves are mapped back to the original sampling points (to match the dimension of the aligned raw data) by linear interpolation.

In order to fit our genetic FMEM, we also need the relationship matrix $\bm{A}$, which can be easily computed by the R package \textbf{pedigreemm} \citep{vazquez_2010_technical}.

\subsection{Results of the aligned model}
We illustrate the estimated covariance functions and genetic eigenfunctions. Simultaneous confidence bands for estimates are constructed using 300 bootstrap samples (singular-fitted samples removed). 

\begin{figure}[h]
     \centering
    \includegraphics[width=\textwidth]{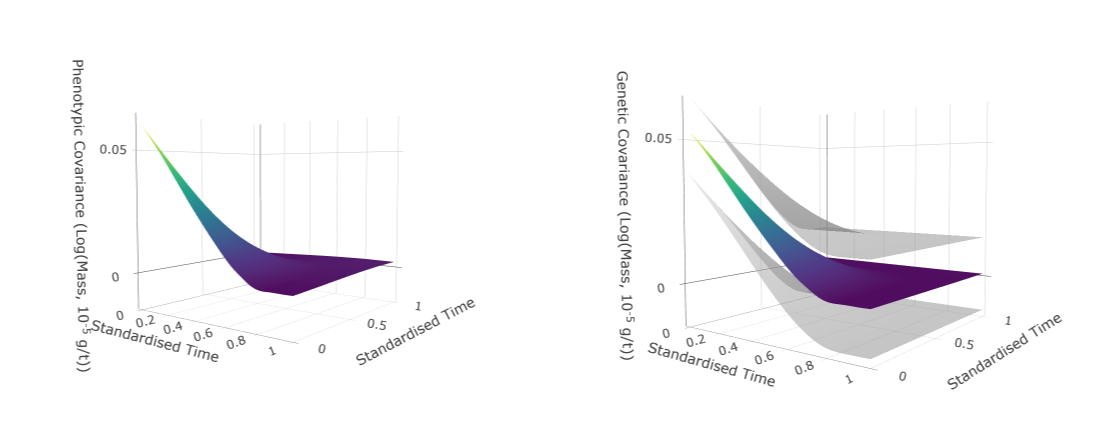}
    \label{fig:TCalignedSCBGCOV}
     \caption{Plot of the estimated phenotypic covariance function (left) and additive genetic covariance function (right) on a logarithmic scale. The grey surfaces are 95\% quantile boundaries.}
     \label{fig:TCalignedcov}
\end{figure}

The phenotypic (left) and additive genetic function (right) are displayed in Figure~\ref{fig:TCalignedcov}. Both functions exhibit a similar shape and scale, suggesting that a significant part of the phenotypic variance is attributed to the additive genetic effect. This finding is consistent with the results in \citet{irwin_2013_constraints} regarding the decomposition of phenotypic variation.

Variation patterns in the phenotypic and additive genetic functions suggest that body mass at an early age is highly correlated with mass at other early ages, but this correlation weakens over time. Surfaces representing variance within the covariance functions imply that additive genetic and phenotypic variations are greatest at early ages and decrease as time progresses.

\begin{figure}[h]
    \centering
    \includegraphics[width=0.5\textwidth]{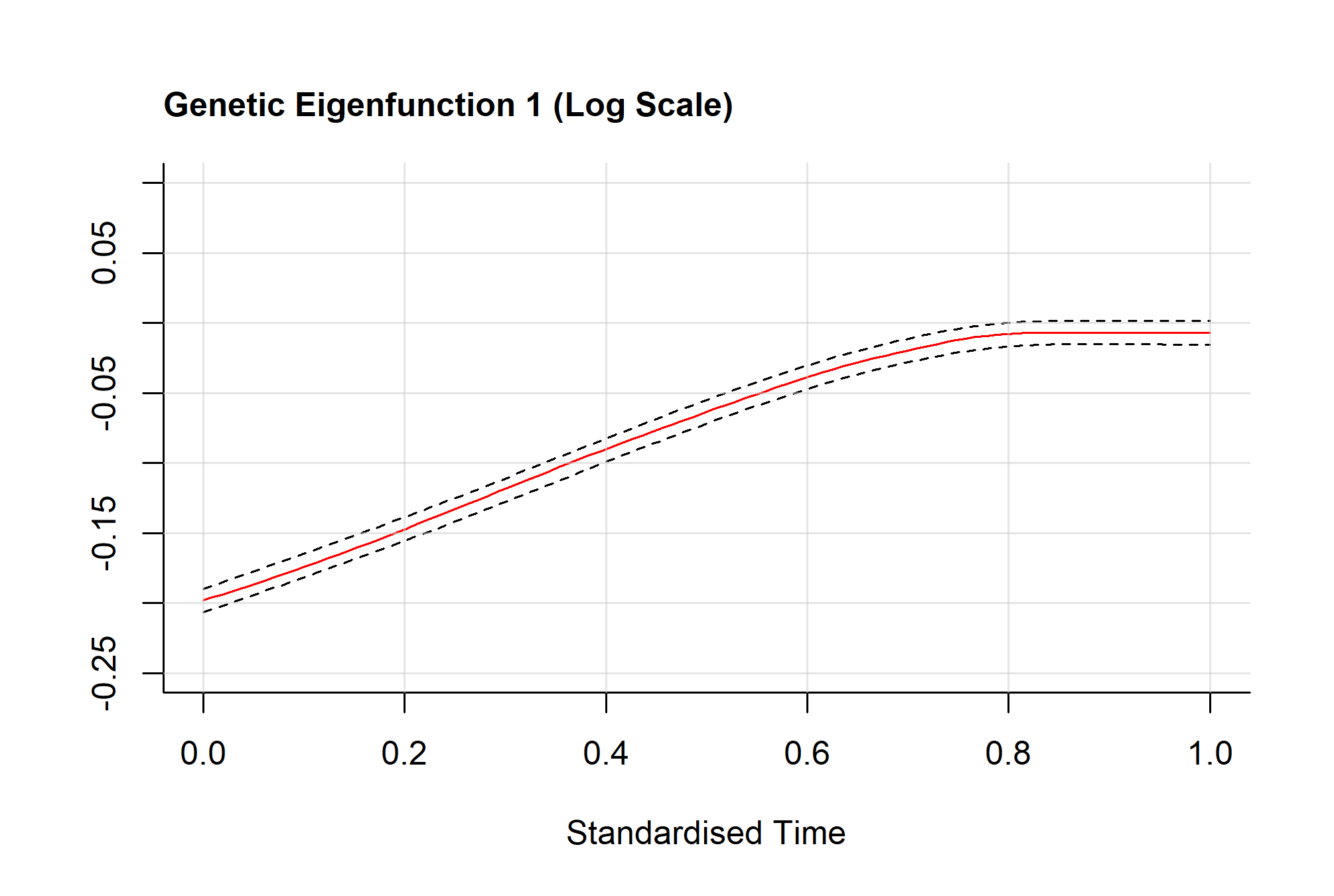}
    \caption{Plot of the estimated genetic eigenfunction (red solid line). The black dashed lines are 95\% quantile boundaries.}
    \label{fig:TCSCBGE}
\end{figure}

As explained in Section~\ref{Sec:GFMEM}, the maximum response to selection corresponds to the direction of the leading genetic eigenfunction. The first eigenfunction, which explains over 93\% of the genetic variation, is shown in Figure~\ref{fig:TCSCBGE}. It starts at a negative value of approximately -0.25 and gradually increases to level off near zero. This eigenfunction captures the overall growth trend, which implies that early differences in growth diminish over time, leading to a more stabilised body mass as individuals approach maturity.

\subsubsection{The impact of time alignment}
\begin{figure}[h]
     \centering
     \begin{subfigure}[h]{0.325\textwidth}
         \centering
         \includegraphics[width=\textwidth]{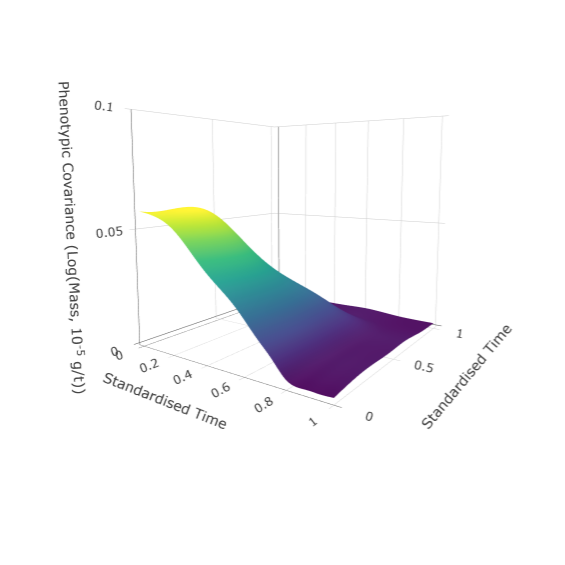}
         \label{fig:TCmisalignedP}
     \end{subfigure}
     \begin{subfigure}[h]{0.325\textwidth}
         \centering
         \includegraphics[width=\textwidth]{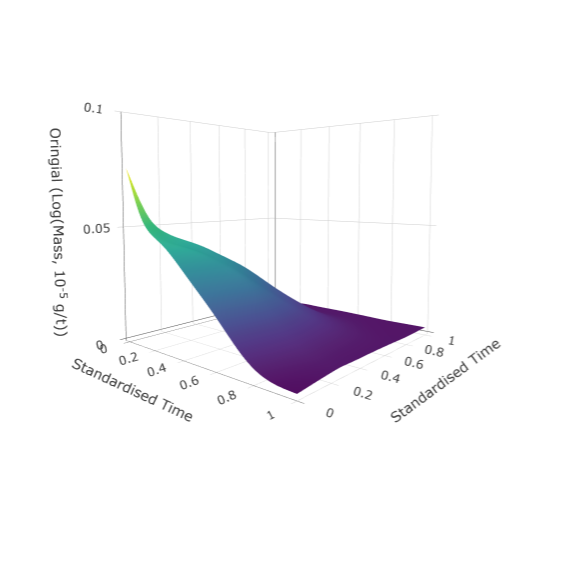}
         \label{fig:TCsampleCov}
     \end{subfigure}
     \begin{subfigure}[h]{0.325\textwidth}
         \centering
         \includegraphics[width=\textwidth]{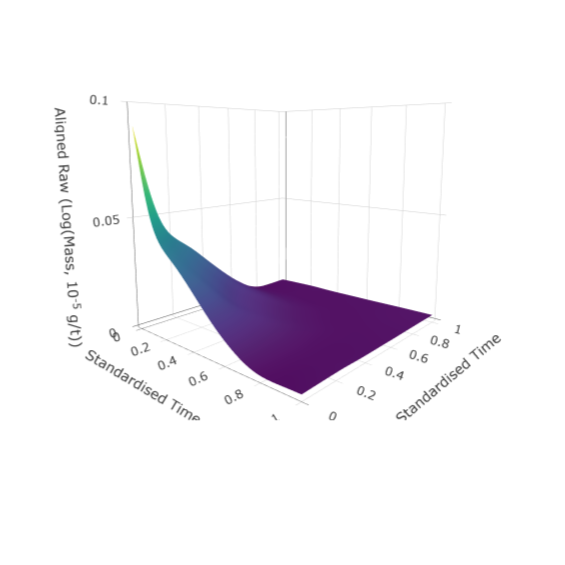}
         \label{fig:TCalignedCov}
     \end{subfigure}
     \caption{Plot of the estimated phenotypic covariance function of original data (left), sample covariance function of the original data (middle) and sample covariance function of the aligned data (right) on a logarithmic scale.}
     \label{fig:TCsamcov}
\end{figure}

To assess the impact of time warping, we also fit the original data with FPCs as the functional basis. This is different from the analysis in \citet{irwin_2013_constraints}, where polynomial basis functions were used. The phenotypic covariance function for unregistered data (Figure~\ref{fig:TCsamcov}, left) is compared with the sample covariance functions for both the original data (Figure~\ref{fig:TCsamcov}, middle) and the aligned data (Figure~\ref{fig:TCsamcov}, right). The phenotypic and sample covariances without alignment look inflated between $0.2 < t < 0.6$. In contrast, the aligned data show most of the variability to be concentrated at the beginning with a consistently decreasing trend. This result illustrates how phase variability interferes with amplitude variability, and this can be problematic when this variability is interpreted only in terms of amplitude variation. We claim that if the phase variability is of interest, this needs to be modelled explicitly, as in the joint model we described in Section~\ref{Sec:Joint}. 

As mentioned, the original paper \citep{irwin_2013_constraints} estimated the covariance functions using misaligned data with orthogonal Legendre polynomials as basis functions. Their results (\cite{irwin_2013_constraints}, Figure~3) showed negative correlations between the body mass at later ages and the one at earlier ages, with a spike of variability for older ages. \citet{irwin_2013_constraints} recognised these results as potentially problematic, noting that they don't match those for many other insect species and proposing that aligning time and significant biological events, such as pupation, might provide results that are more biologically insightful. In contrast, we do not observe this behaviour in the sample covariance function (Figure~\ref{fig:TCsamcov}, right), and it seems natural to investigate if this may be due to time misalignment. For a proper comparison, we also fit our model using Legendre polynomials without registering nor rescaling the curves on the same interval. In this case, we observe the same results reported in \citet{irwin_2013_constraints} for the covariance functions on the original time scale. On the other hand, if we scale curves with common starting and ending times, the estimated covariance functions are similar to the results obtained using FPCs as basis functions. This suggests that the results are robust to the choice of the basis functions, but indeed, as \citet{irwin_2013_constraints} suspected, the time misalignment may lead to difficulties in the interpretation of the results of the analysis.



\subsection{Joint analysis of amplitude and phase variations}
We fit the growth data to the joint FMEM introduced in Section~\ref{Subsec:JointFMEM}. The same data pre-processing is applied: first, standardise time domains; then, smooth and register curves to obtain amplitude and warping functions. We smooth data on a fine time grid to better capture shapes of growth trajectories. This detailed representation allows the shape-preserving registration by the Fisher-Rao metric to compute a sufficient number of distinct warping functions, which align features in the data while maintaining the overall data structure. We use the CLR transformation on the derivative of warping functions to get the unconstrained phase functions. To improve computational efficiency, amplitude and phase functions are interpolated to a time grid of eight points since the original samples have, on average, eight measurements per individual.

\begin{figure}[h]
    \centering
    \includegraphics[width=\linewidth]{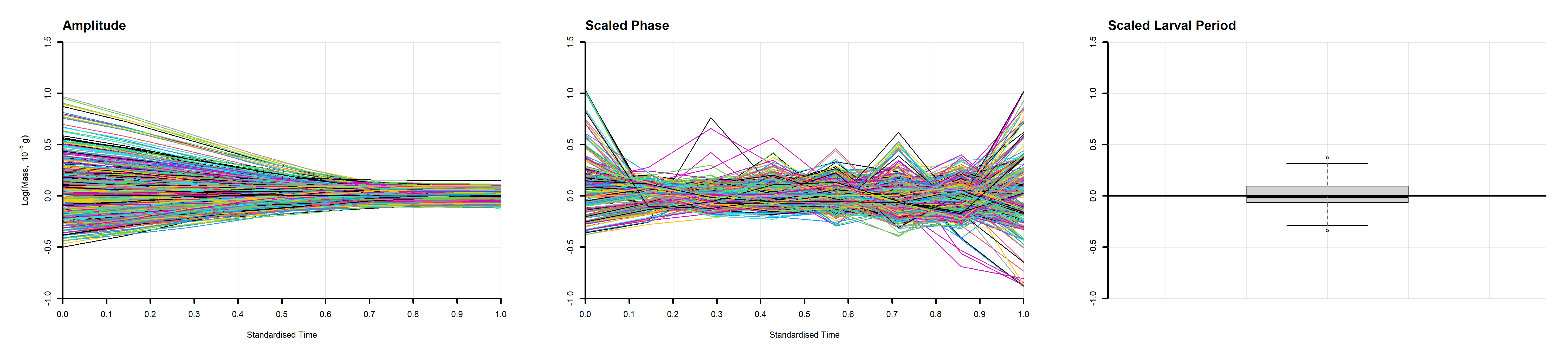}
    \caption{Plot of the joint curves: the centred amplitude functions (left), the scaled phase functions (middle) and the scaled individual larval periods (right).}
     \label{fig:jointcurves}
\end{figure}

We combine the amplitude functions, phase functions and the individual larval period in the form of Equation~(\ref{eqn:gc2}). Since we standardise time domains in the first step, we include the larval period to account for individual variability in duration, which is not explicitly captured in the amplitude or phase functions. However, the scale of the larval period is not in the same order as the amplitude functions. Therefore, we rescale the larval period length with respect to the variability of the amplitude functions. We also observe that the phase functions vary around zero. The amplitude functions are centred with mean zero to ensure no sudden jump between the amplitude and phase functions. Finally, we choose the scaling parameter C, which glues the amplitude functions and the phase functions using Equation~(\ref{eqn:computeC}). We use the maximum norm, i.e., $\norm{A}_{\text{max}} = \max_{i,j}\absnorm{a_{ij}}$, to match the maximum variability in both amplitude and phase components. Figure~\ref{fig:jointcurves} displays the joint functions in all three components.

\begin{figure}[h]
    \centering
    \includegraphics[width=\linewidth]{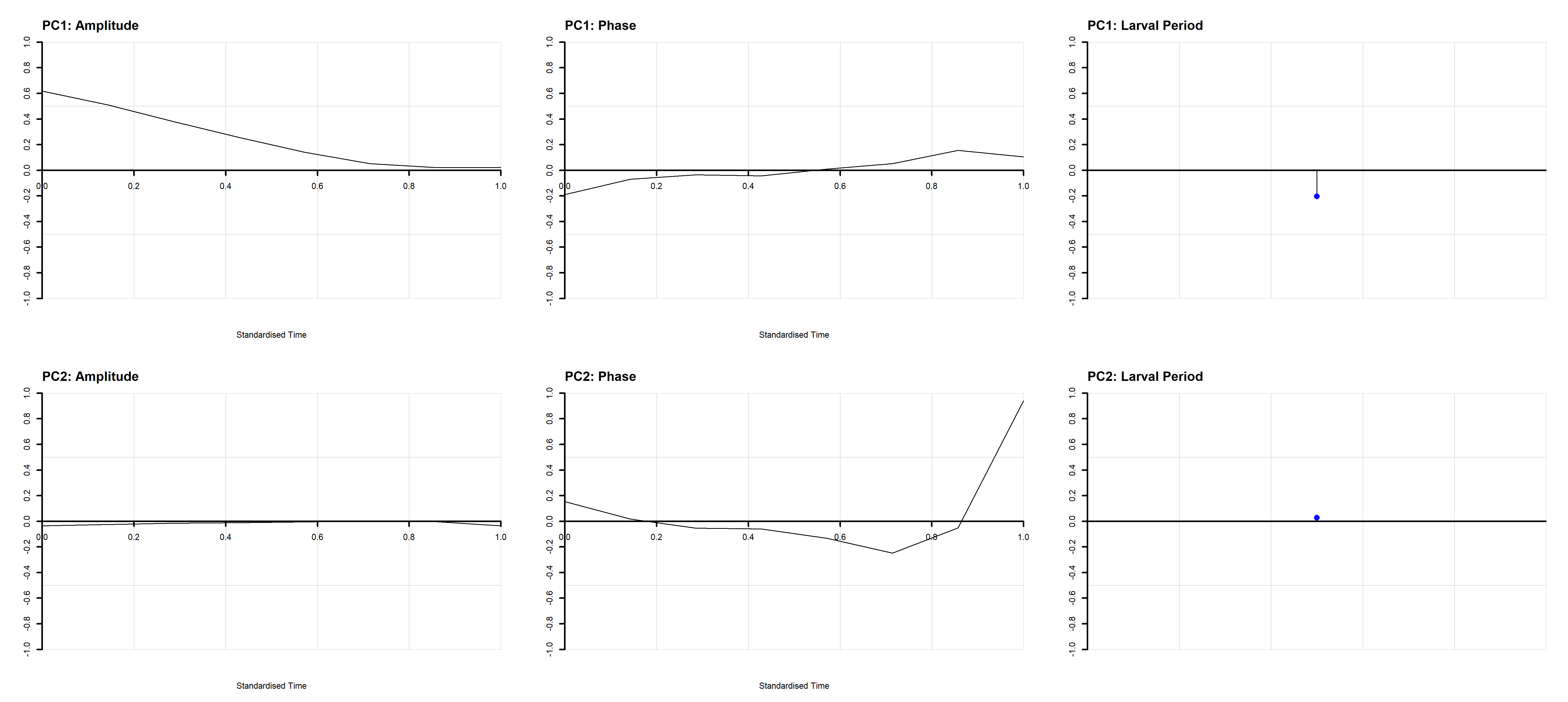}
    \caption{Plots of the first (top row) and second (bottom row) combined FPCs. Left: the amplitude FPCs. Middle: the phase FPCs. Right: the larval period PCs.}
     \label{fig:combinedFPCs}
\end{figure}

We run FPCA on the joint curves. The combined FPCs have three parts. The first and second functional parts describe the dominant variational pattern in the amplitude and phase functions, respectively. The third scalar part explains how the length of the larval period is associated with the growth pattern. We show the first two combined FPCs in Figure~\ref{fig:combinedFPCs}. The first combined FPC primarily captures the variation in the amplitude functions and is virtually identical to the first FPC extracted from the amplitude functions alone. Meanwhile, the phase effect contributes more to the second combined FPC. Before we fit the joint model, we need to determine the number of combined FPCs to use. To balance the amount of variational trends retained and the computational time, we use five combined FPCs, explaining 89\% total variance, as basis functions for both random components and we use six combined FPCs for the mean shape.

\subsubsection{Results of the joint model}

\textbf{Estimation of covariance functions}

We first show the estimated covariance functions. Figure~ref{fig:TCcombinedampcov} - Figure~\ref{fig:TCcombinedxcov} display the covariance of amplitude functions, phase functions and cross-covariance between them, respectively. In each case, we compare the estimate using sample data with the phenotypic covariance estimated from the joint model (shown in the top row). A valid phenotypic covariance structure is expected to align with the sample estimate. The corresponding additive genetic and environmental components are illustrated in the bottom row. 
\begin{figure}[h]
    \centering
    \includegraphics[width=0.8\linewidth]{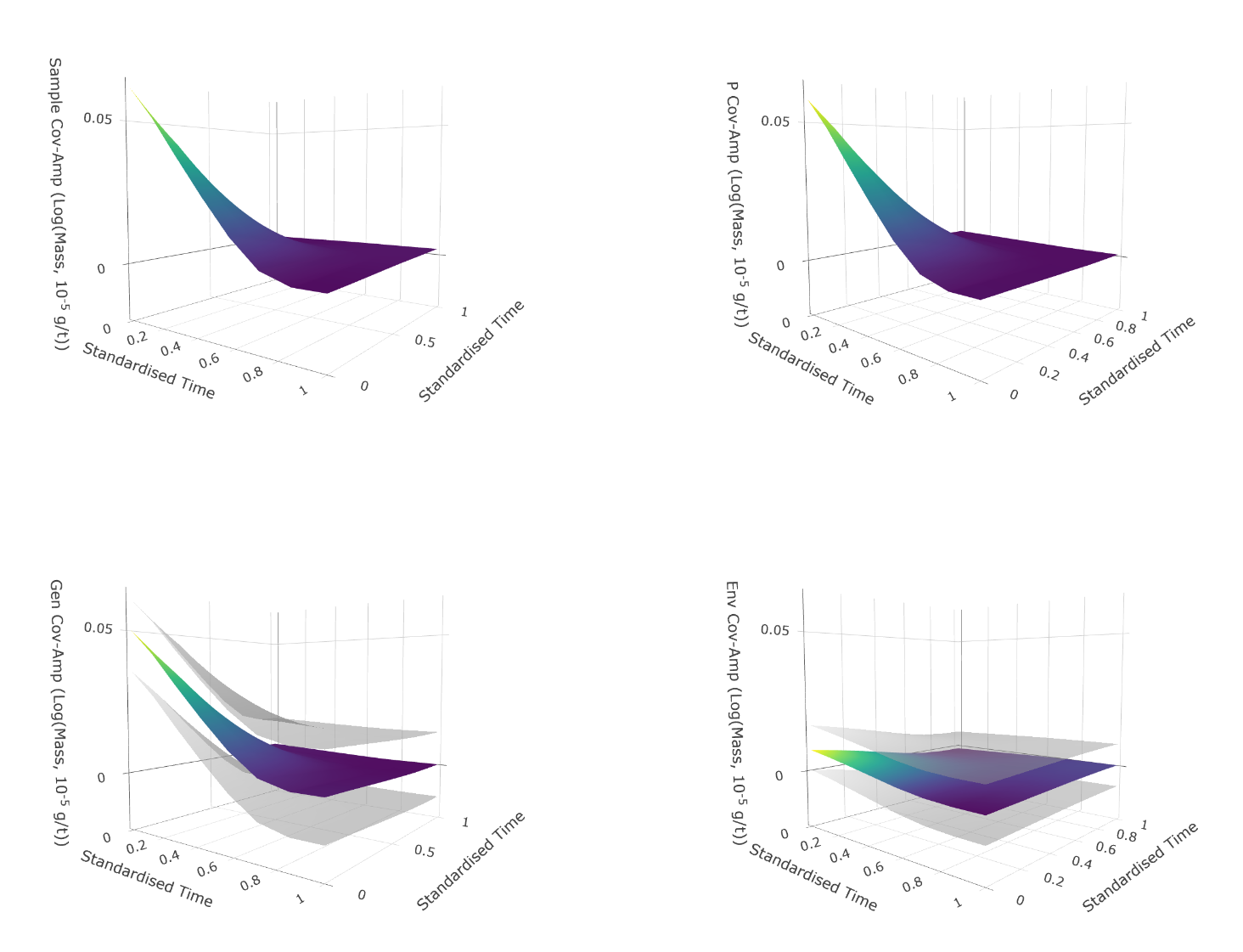}
    \caption{Results of the amplitude covariance function. Top left: sample amplitude covariance. Top right: Phenotypic amplitude covariance. Bottom left: genetic amplitude covariance. Bottom right: environmental amplitude covariance. The grey surfaces represent 95\% quantile boundaries.}
     \label{fig:TCcombinedampcov}
\end{figure}

\begin{figure}[h]
    \centering
    \includegraphics[width=0.8\linewidth]{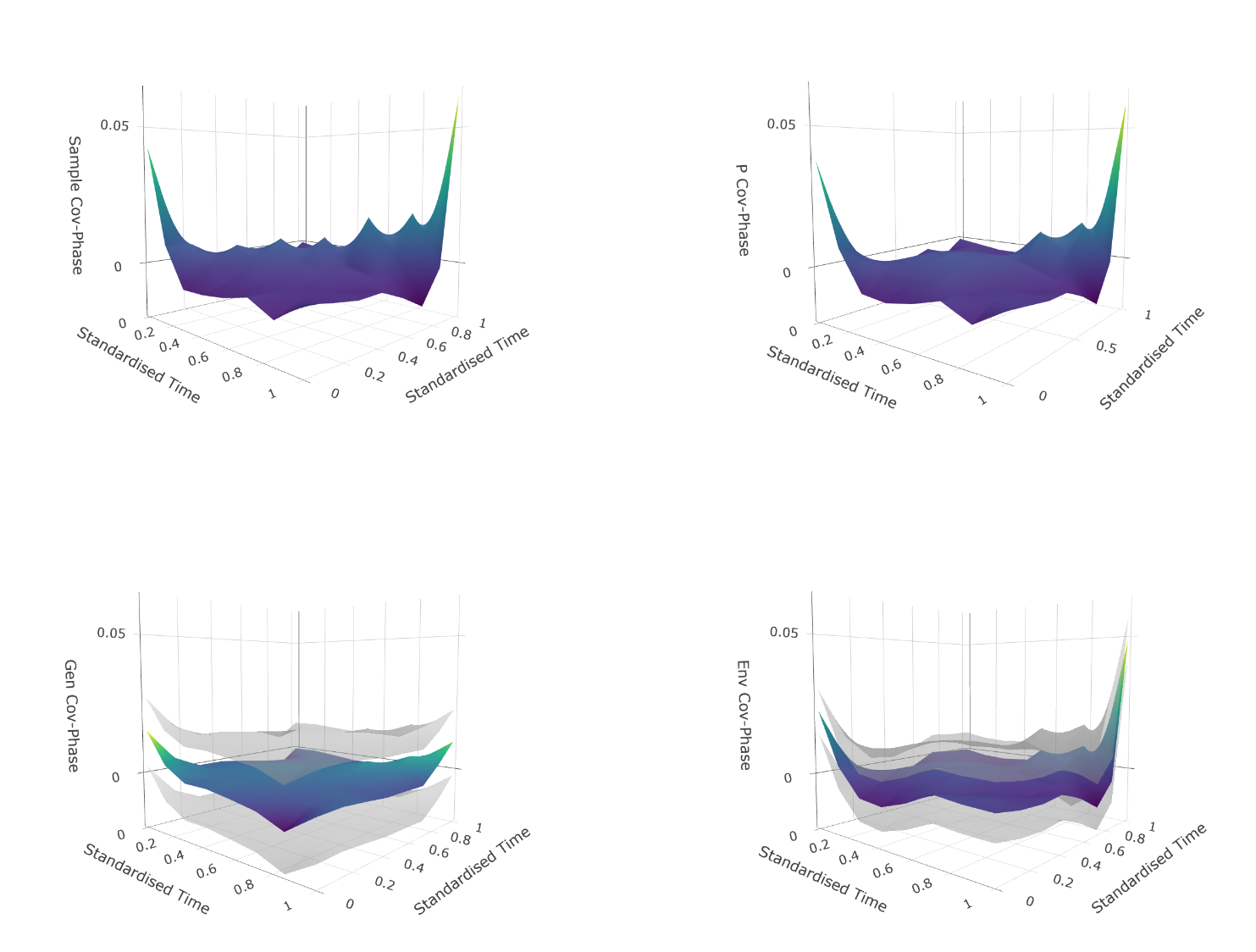}
    \caption{Results of the phase covariance function. Top left: sample phase covariance. Top right: Phenotypic phase covariance. Bottom left: genetic phase covariance. Bottom right: environmental phase covariance. The grey surfaces represent 95\% quantile boundaries.}
     \label{fig:TCcombinedphase}
\end{figure}

\begin{figure}[h]
    \centering
    \includegraphics[width=0.8\linewidth]{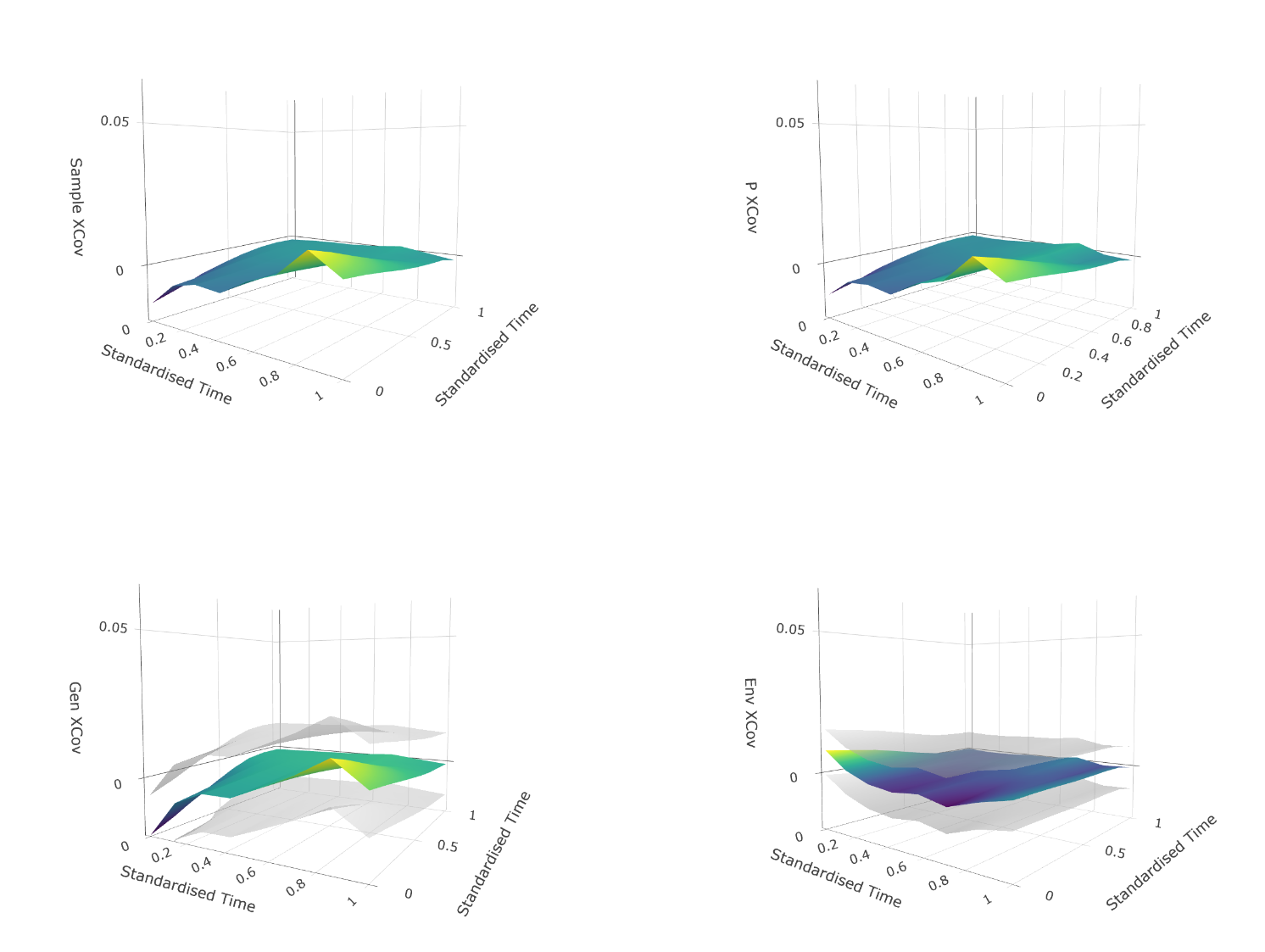}
    \caption{Results of the cross-covariance function between amplitude and phase. Top left: sample cross-covariance. Top right: Phenotypic cross-covariance. Bottom left: genetic cross-covariance. Bottom right: environmental cross-covariance. The grey surfaces represent 95\% quantile boundaries.}
     \label{fig:TCcombinedxcov}
\end{figure}

\begin{figure}[h]
    \centering
    \includegraphics[width=0.9\linewidth]{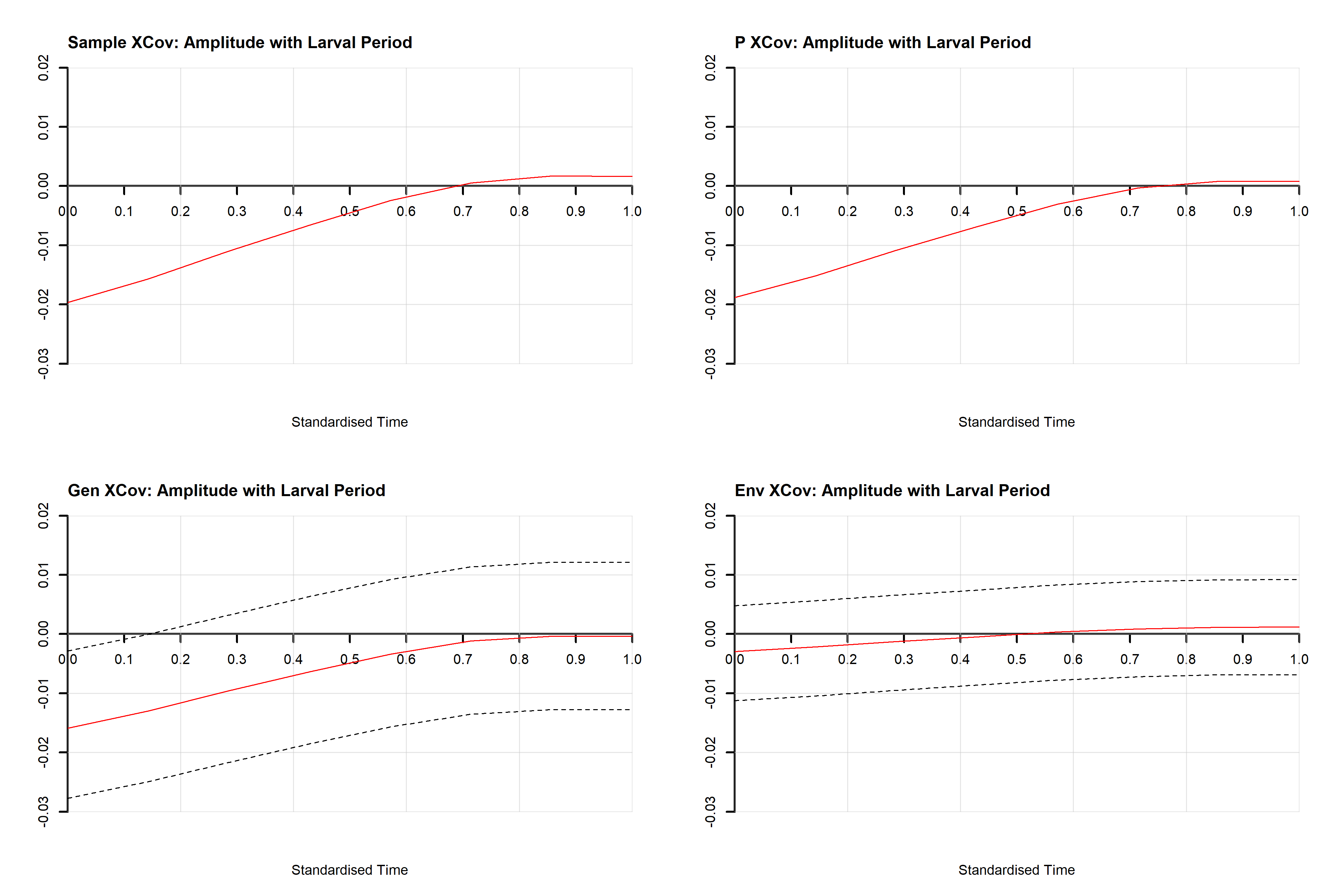}
    \caption{Results of the cross-covariance function between amplitude and larval period. Top left: sample cross-covariance. Top right: Phenotypic cross-covariance. Bottom left: genetic cross-covariance. Bottom right: environmental cross-covariance. The black dashed lines are 95\% quantile boundaries.}
     \label{fig:amplarxcov}
\end{figure}

\begin{figure}[h]
    \centering
    \includegraphics[width=0.9\linewidth]{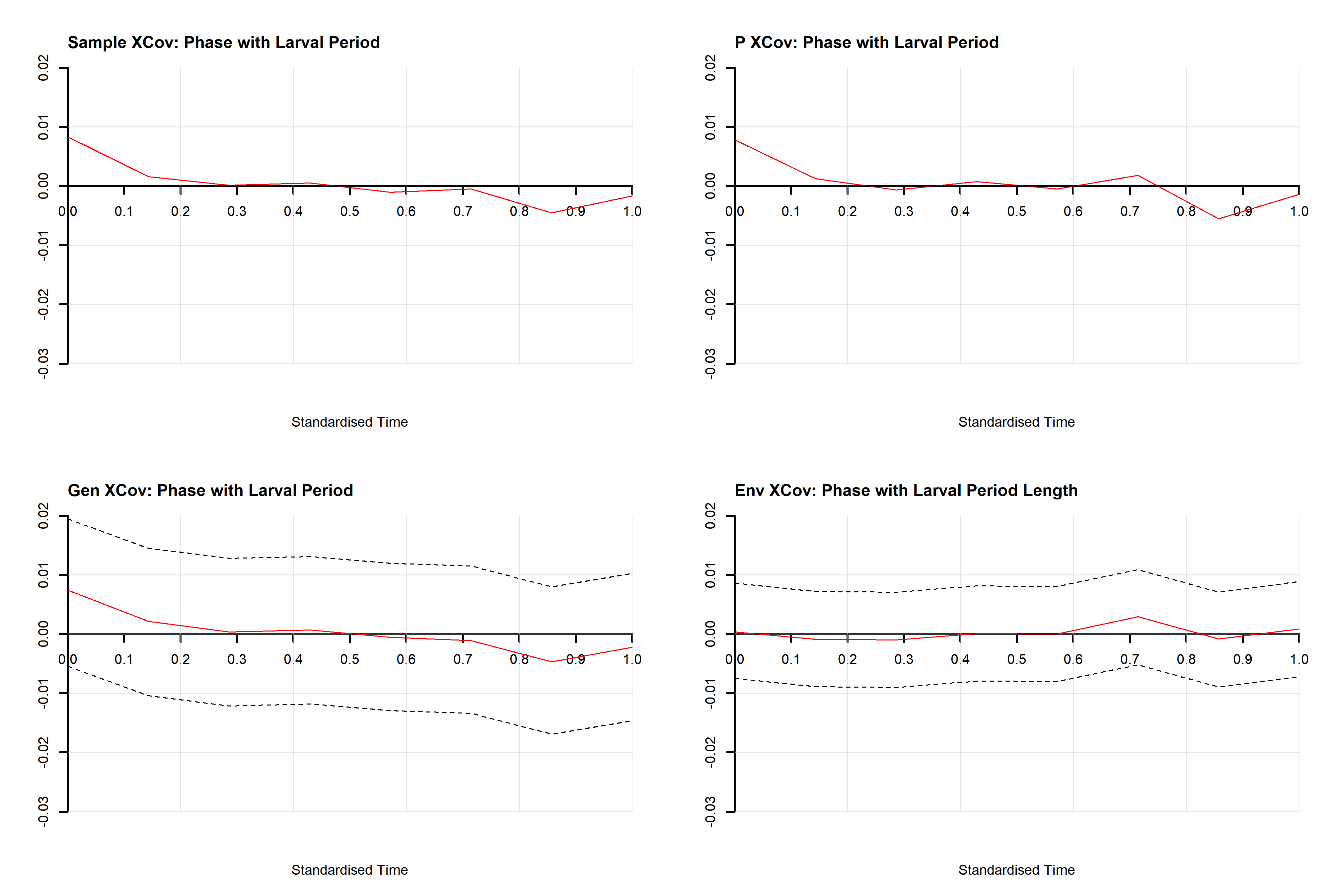}
    \caption{Results of the cross-covariance function between phase and larval period. Top left: sample cross-covariance. Top right: Phenotypic cross-covariance. Bottom left: genetic cross-covariance. Bottom right: environmental cross-covariance. The black dashed lines are 95\% quantile boundaries.}
     \label{fig:phaselarxcov}
\end{figure}

\begin{figure}[h]
    \centering
    \includegraphics[width=0.9\linewidth]{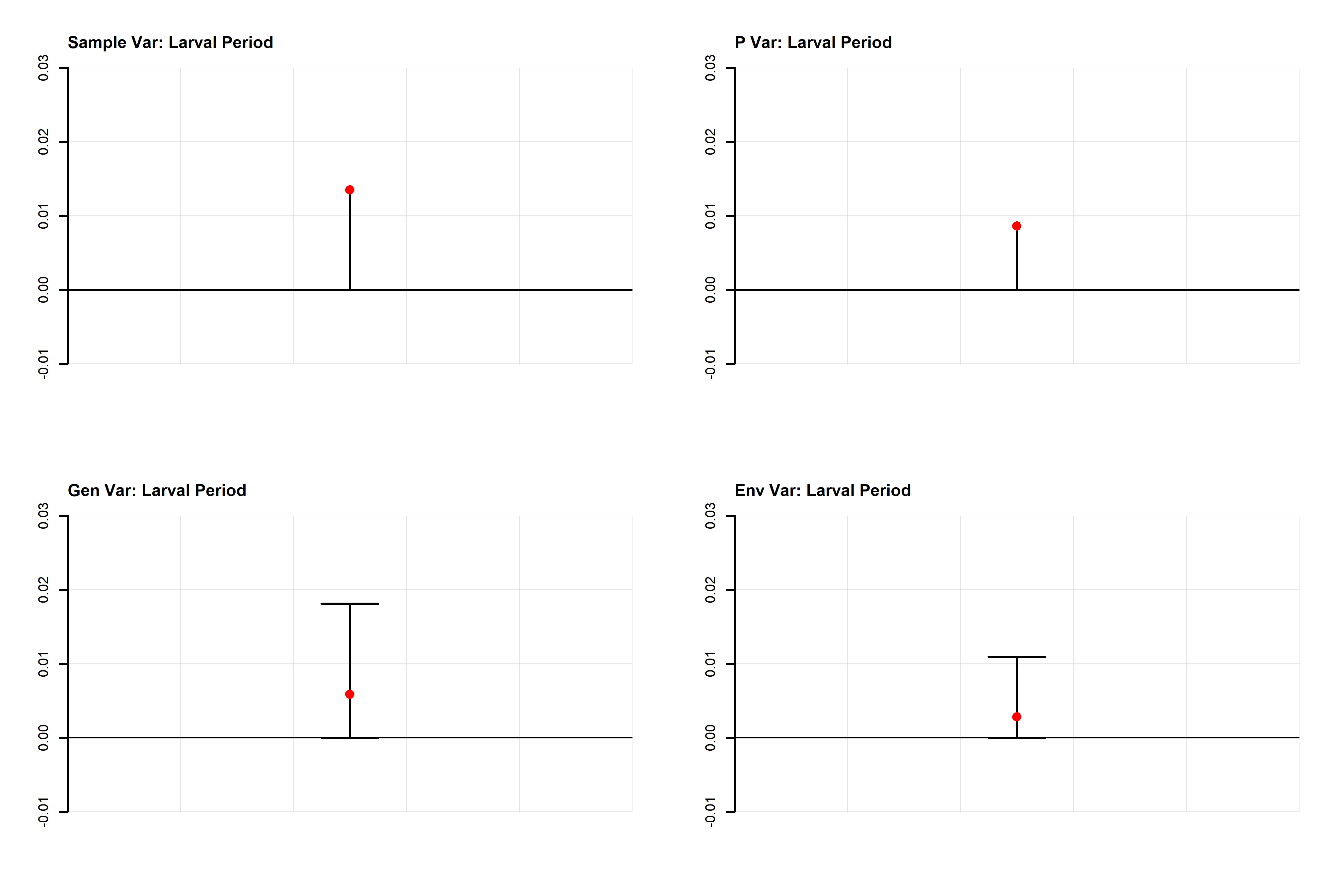}
    \caption{Results of the variance of the larval period. Top left: sample component. Top right: Phenotypic component. Bottom left: genetic component. Bottom right: environmental component.}
     \label{fig:larvar}
\end{figure}
Focusing on the amplitude covariance function, we see the phenotypic covariance is visually identical to the result of the aligned model, demonstrating that the additive genetic effect dominates the phenotypic variation. Moving to the phase covariance function, the most significant variation occurs along the diagonal, suggesting stronger local correlations. Comparing its genetic and environmental components, the environmental effect seems to strongly influence the phase variation. The cross-covariance between the amplitude and phase is weak, and the interpretation is ambiguous.

Figure~\ref{fig:amplarxcov} plots the cross-covariance between the amplitude and larval period. An initially negative then positive cross-covariance suggests that an individual with a longer larval period may have a smaller initial mass but experience a faster increase at a later stage. This growth pattern could be explained by the environmental influence, as the environmental cross-covariance is positive at a later age, whereas the genetic component is negative all the time. Next, we show the cross-covariance between the phase and larval period in Figure~\ref{fig:phaselarxcov}. We observe the opposite trend with initial positive then negative covariance, and the additive genetic effect dominates the covariation pattern. However, all values are close to zero, so interpreting these patterns should be done with caution.

The variance of the larval period is shown in Figure~\ref{fig:larvar}. It suggests that the variance explained by the additive genetic component is nearly twice that of the environmental part.

\clearpage
\textbf{Selection gradient}

\begin{figure}[h]
    \centering
    \includegraphics[width=\linewidth]{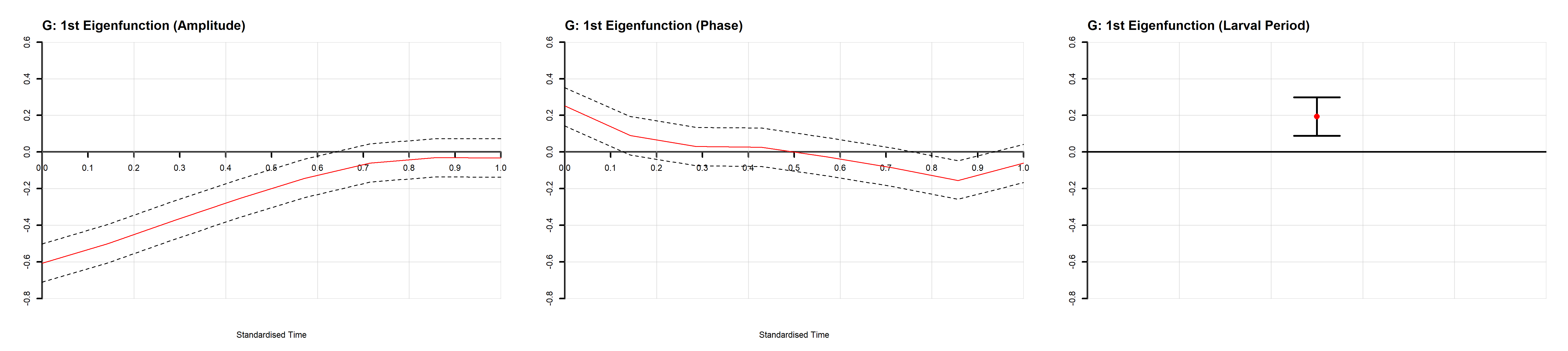}
    \caption{Plot of the first genetic eigenfunction. The black dashed lines are 95\% quantile boundaries. }
     \label{fig:scbgeneigen}
\end{figure}

\begin{figure}[h]
     \centering
     \begin{subfigure}[h]{0.495\textwidth}
         \centering
         \includegraphics[width=\textwidth]{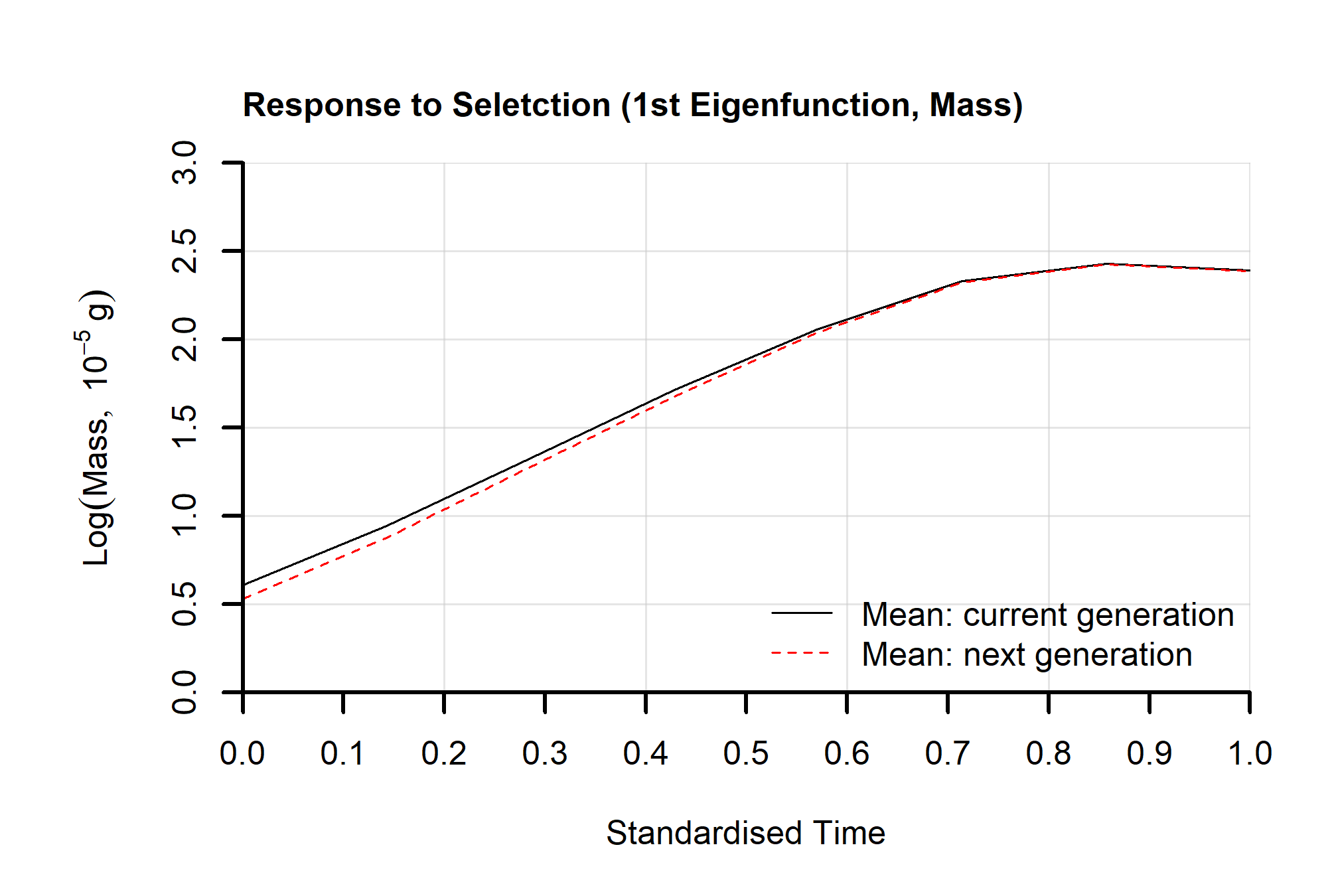}
         \caption{}
         \label{fig:combinedevo}
     \end{subfigure}
     \begin{subfigure}[h]{0.495\textwidth}
         \centering
         \includegraphics[width=\textwidth]{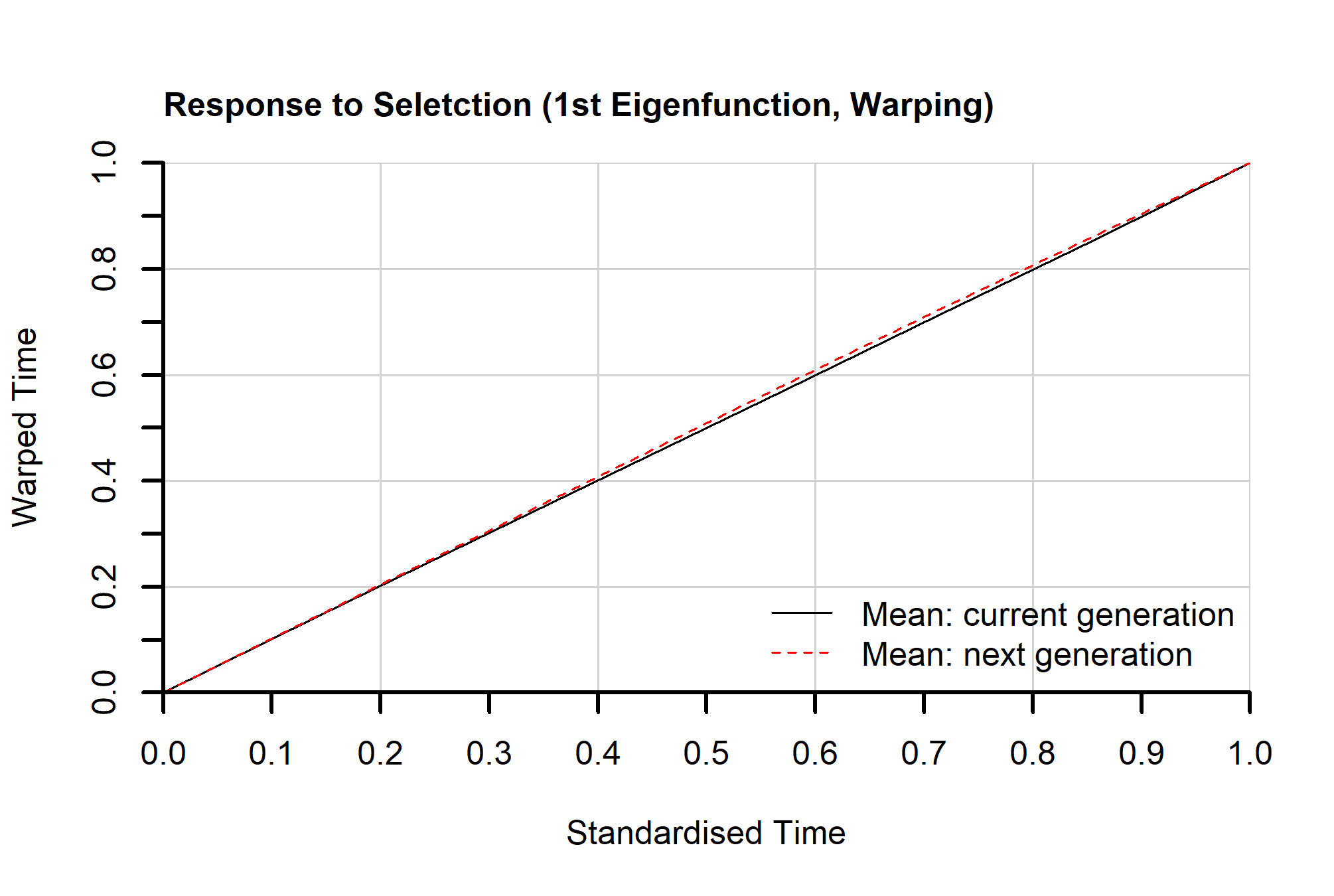}
         \caption{}
         \label{fig:combinedwarp}
     \end{subfigure}
     \caption{(a) Evolutionary response to selection of the amplitude function in the direction of the first genetic eigenfunction. (b) Evolutionary response to selection of the warping function in the direction of the first genetic eigenfunction.}
     \label{fig:combinedselection}
\end{figure}

Figure~\ref{fig:scbgeneigen} shows the estimated leading genetic eigenfunction. We demonstrate the evolutionary response to hypothesised selection (in the direction of the leading genetic eigenfunction) in the mass (amplitude variation) in Figure~\ref{fig:combinedselection} (a) and the warping (phase variation) in Figure~\ref{fig:combinedselection} (b).  The change in mean mass is similar to the result of the aligned model; individuals with a smaller initial mass may experience accelerated growth in the middle stage until body mass stabilises before maturity. This trend is reflected in the inter-generation change in the mean warping function, where a slight increase in the middle stage is observed.

%% file: 6_Discussion.tex
\section{Discussion and future directions}\label{Sec:Discussion}
We have proposed a genetic FMEM to estimate the covariance functions of FV traits. First, functional data have been registered in the data pre-processing stage using shape-preserving registration based on the Fisher-Rao metric \citep{srivastava_2011_registration}. Then, a functional mixed-effects model has been fitted using a functional principal component basis, since data-driven basis functions provide an optimal framework for capturing the underlying structure and variability in the observed functional data. For numerical computation, we use \textbf{lme4} with model reparameterisation to analyse genetic data in a functional framework. We demonstrated in simulations that the estimation procedure recovers the true parameters and that simultaneous bootstrap confidence bands are able to capture the uncertainty.

Although registering curves in the pre-processing stage is a straightforward solution to curve misalignment, it disregards all phase variations. We therefore introduced a joint model which simultaneously analyses amplitude functions, warping functions and durations of curves via a combined (functional) PCA  approach. 

The aligned and joint models are applied to a real-case study of \textit{Tribolium castaneum} dataset, with a particular interest in estimating the additive genetic covariance function and its principal eigenfunctions. The eigenfunctions, which are also equivalent to the selection gradient function that would produce the maximal evolutionary response, play a significant role in predicting evolutionary response to selection and identifying evolutionary constraints. Our estimation from both models implies that the genetic effect dominates the phenotypic variation, which is consistent with the results of \citet{irwin_2013_constraints}. However, they found a negative correlation in body mass at later ages compared to earlier times, while we do not observe this negative correlation. We suggest that this difference is caused by scaling time proportionally to the pupation time (larval period), which also highlights the need to incorporate the duration of curves into the joint model. Focusing on the results of the aligned model, phase variations are primarily explained by the environmental effect.  Despite the challenges that exist in the interpretation of warping effects in the genetic framework through the transformed phase functions, this can offer greater biological explanatory power than simply focusing on the aligned curves.

One limitation of our methods is the numerical instabilities that can lead to singular fit for the model, where the covariance matrix is less than full rank. Our current solution is to use a smaller number of FPCs to avoid overfitting the model. However, a drawback of this approach is the lack of flexibility in determining the number of basis functions. Therefore, the first direction of future work is to look for robust estimation methods for the genetic mixed-effects models, which could allow us to use more flexible basis functions.

Another important direction for future research is the integration of amplitude and warping in mixed-effects models. Instead of analysing the (transformed) phase variations via FPCA, warping can be considered as a nonlinear random effect. The nonlinearity due to the presence of warping functions must be carefully addressed. \citet{larslauraket_2013_a} proposed a functional mixed-effects model where the misalignment is introduced in the common mean. Estimation was done via a linearisation of the mean around the warped values. A different approach is the two-level, nonlinear functional mixed-effects model using B-spline representation discussed by \citet{claeskens_2021_nonlinear}. They used a linear structure to deal with individual amplitude variations and a nonlinear model in the warping functions to account for individual phase variations. Both methods investigate general warped mixed-effects models and could be modified to analyse genetic data.